# PROPERTIES OF THE BARE NUCLEUS OF COMET 96P/MACHHOLZ 1 *

Nora L. Eisner[1,2,3], Matthew M. Knight[2], Colin Snodgrass[4,5], Michael S. P. Kelley[2], Alan Fitzsimmons[6]
and Rosita Kokotanekova[5,7,8]

[1] Contacting author: *nora.eisner@physics.ox.ac.uk*
[2] Department of Astronomy, University of Maryland, College Park, MD 20742, USA
[3] Department of Astrophysics, University of Oxford, Oxford, UK
[4] Institute for Astronomy, University of Edinburgh, Royal Observatory, Edinburgh EH9 3HJ, UK
[5] Planetary and Space Sciences, School of Physical Sciences, The Open University, Milton Keynes, MK7 6AA, UK
[6] Astrophysics Research Centre, Queen's University Belfast, Belfast BT7 1NN, UK
[7] European Southern Observatory, Karl-Schwarzschild-Strasse 2, 85748 Garching bei München, Germany
[8] Max-Planck-Institut für Sonnensystemforschung, Justus-von-Liebig-Weg 3, 37077 Göttingen, Germany

## ABSTRACT

We observed comet 96P/Machholz 1 on a total of 9 nights before and after perihelion during its 2017/2018 apparition. Both its unusually small perihelion distance and the observed fragmentation during multiple apparitions make 96P an object of great interest. Our observations show no evidence of a detectable dust coma, implying that we are observing a bare nucleus at distances ranging from 2.3 AU to 3.8 AU. Based on this assumption we calculated its color, and found average values of g'-r' = 0.50 ± 0.04, r'-i' = 0.17 ± 0.03, and i'-z' = 0.06 ± 0.04. These are notably more blue than those of the nuclei of other Jupiter family and long period comets. Furthermore, assuming a bare nucleus, we found an equivalent nuclear radius of 3.4 ± 0.2 km with an axial ratio of at least 1.6 ± 0.1. The lightcurve clearly displays one large peak, one broad flat peak, and two distinct troughs, with a clear asymmetry that suggests that the shape of the nucleus deviates from that of a simple triaxial ellipsoid. This asymmetry in the lightcurve allowed us to constrain the nuclear rotation period to 4.10 ± 0.03 hours and 4.096 ± 0.002 hours before and after perihelion, respectively. Within the uncertainties, 96P's rotation period does not appear to have changed throughout the apparition, and we conclude a maximum possible change in rotation period of 130 seconds. The observed properties were compared to those of comet 322P and interstellar object 1I/'Oumuamua in an attempt to study the effects of close perihelion passages on cometary surfaces and their internal structure, and the potential interstellar origin of 96P.

## 1. INTRODUCTION

In this paper, we focus on analyzing the properties of the unusual comet 96P/Machholz 1 (henceforth 96P). Although 96P has a short orbital period (~5.3 years), its orbit is highly inclined (~58°) and it has a very low perihelion distance ($q$; currently 0.124 AU). These properties are very different from typical Jupiter family comet orbits. 96P has been shown to be trapped in a Kozai resonance with Jupiter (e.g., Levison & Dones 2014), where its eccentricity and inclination oscillate out of phase but with the same frequency. Orbital integrations (e.g., Green et al. 1990, McIntosh 1990) find that it oscillates from a high-inclination orbit (near 80°) with perihelion near 1 AU, to a low-inclination orbit with perihelion at ~0.03 AU, and back again on a timescale of ~4000 years. Although 96P's Tisserand parameter ($T_J$) is currently 1.94, Bailey et al. (1992) showed that $T_J$ was greater than 2 a few thousand years ago and has been slowly drifting downward. Given that $T_J$ = 2.0 is the canonical divider (e.g., Carusi et al. 1987) between Jupiter family orbits (2.0 < $T_J$ < 3.0) and Halley-type or Oort cloud orbits ($T_J$ < 2.0), 96P's origin is uncertain.

Comet 96P is generally difficult to study; it has a small solar elongation when near perihelion, it never comes particularly close to Earth due to its high inclination, and its high southern declination throughout most of its orbit limits the telescopes capable of observing it. It has been observed on every apparition since 1996 by the space-based *SOlar and Heliospheric Observatory* (*SOHO*)'s SWAN UV imager (Combi et al. 2011, 2019) and LASCO optical imager (Grynko et al. 2004, Eisner et al. 2018). However, ground-based studies have been relatively limited. Near-aphelion observations by Meech (1996) and Licandro et al. (2000) indicated that it has a large (equivalent radius of 3.2 km), elongated (axial ratio > 1.4) nucleus and suggested a rotation period of 6.38 hours (see also Lamy et al. 2004). Sekanina (1990) analyzed various publicly available datasets and concluded that it has very low activity for its size, suggesting that this likely hampered an earlier discovery. Most intriguingly, Langland-Shula and Smith (2007) and Schleicher (2008) found it to be depleted in $C_2$ and $C_3$ and highly depleted in CN, with Schleicher (2008) suggesting this combination is primordial and possibly due to an interstellar origin.

The current work is motivated by our interest in the properties of objects with low perihelion distances. Despite the aforementioned challenging viewing geometry, 96P is the most accessible comet on a low-$q$ orbit, potentially allowing evolutionary effects to be detected on short timescales. Its perihelion distance is less than half that of 2P/Encke's ($q = 0.336$ AU), the comet typically most associated with evolutionary effects (e.g., Sekanina 1988, 1991). The only short period comet with a perihelion distance smaller than 96P's to be observed from the ground is 322P/SOHO 1 ($q = 0.054$ AU), but 322P is much smaller and intrinsically fainter. Knight et al. (2016) found that 322P has unusual properties: it was inactive near 1 AU despite behavior consistent with having a coma in the *SOHO* fields of view (Lamy et al. 2013), has a fast rotation period and high albedo suggestive of an asteroidal rather than cometary origin, and has atypical nucleus colors for comet nuclei. Since 96P is unquestionably cometary, comparison of it with 322P may yield insight into whether 322P's properties are evolutionary or primordial. This will aid our understanding of the low-$q$ asteroid population. Granvik et al. (2016) found a significant deficit of asteroids on small perihelion distance orbits, which they suggested was due to the preferential loss of low-albedo objects via disruption on timescales of less than 250 years.

Although the cause is in doubt, 96P already demonstrates behavior of disruption. Two groups of short-period "sunskirting" comets, the Marsden and Kracht groups, were discovered in *SOHO* LASCO images (see Marsden 2005, Battams & Knight 2017, Jones et al. 2018 and references therein). Their orbits were recognized to be similar to future orbits of 96P, and they have been dynamically linked to 96P, along with several meteor streams, as part of the "Machholz Complex" (Ohtsuka et al. 2003; Sekanina & Chodas 2005). The Machholz Complex appears to be continuing to evolve, with some Kracht and Marsden objects failing to be reobserved and new fragments apparently being produced (e.g., Knight 2008).

During its last two perihelion passages, 96P itself has been accompanied by two (in 2012) and three (in 2017) faint fragments when seen in *SOHO* LASCO images (Battams & Lui 2013, Battams[8,9]). Due to the short orbital arcs and resulting large ephemeris uncertainties, it is not certain when these fragments were produced (Sekanina 2013) or even if the same objects were reobserved in 2017, but it is clear that 96P is a prime target for continued monitoring. Despite the small perihelion distance, 96P is still well beyond the Roche limit for a typical cometary density (Knight & Walsh 2013), and tidal forces are not likely to be the primary cause of its ongoing fragmentation (Sekanina & Chodas 2005). Mueller & Samarasinha (2018) argued that it is among the comets most likely to exhibit a large change in rotation period due to torques caused by outgassing. As we will show later, 96P's rotation period is close to the rotational spin limit for a strengthless body so disruption due to rotational spin-up is viable and potentially detectable on short timescales.

We took advantage of 96P's large and relatively inactive nucleus to study it before and after the most recent perihelion passage (2017 October 27.96), allowing us to obtain a high-precision nucleus lightcurve with modest (4 m-class) telescope apertures. We obtained observations on three nights in 2017 and six nights in 2018 (*Section 2*). We did not detect coma or tail and measured nucleus properties such as size and color (*Section 3*) and searched for nearby fragments (*Section 4*). In *Section 5* we present detailed lightcurve analyses, including highly constrained rotation periods pre- and post-perihelion that allow us to look for a change in rotation period. Finally, we discuss and summarize our results in *Sections 6* and *7*, respectively.

## 2. OBSERVATIONS AND REDUCTIONS

### 2.1. *Observing Overview and Reduction*

Images were obtained on a total of nine nights during the 2017/2018 apparition, primarily using the broadband *r'*-filter. The observations were always tracked at the ephemeris rate of the comet, with the exception of the images obtained on 2018 July 11/12 with the New Technology Telescope, which were taken with sidereal tracking and short-enough exposures to prevent the comet from moving more than the seeing disk. A summary of the observations, filters used, and weather conditions is given in *Table 1*. Due to the wide range of telescopes, instruments, and reduction techniques that were implemented, the individual observation runs are described individually in the subsections to follow. Prior to any calibrations, all data were reduced using standard bias and flat-fields techniques. Frames that were significantly contaminated by field stars were discarded. Coherent lightcurves displaying at least one peak and trough were obtained on a total of four nights on 2017 July 2, 2018 July 10/11, 2018 July 31/August 1, and on 2018 August 2. On the remaining five nights of observations we acquired snapshots of comet 96P at one or more epochs during a night.

#### 2.1.1. *2017 April 7*

Images taken on 2017 April 7 were obtained using the 4.2-m Southern Astrophysical Research (SOAR) telescope on Cerro Pachon in Chile using the Goodman Spectrograph Red Camera which utilizes an e2v 231-84 CCD (Clemens et al. 2004). On-chip 2 × 2 binning produced images with a pixel scale of 0.30 arcseconds pixel$^{-1}$. We obtained snapshot observations using SDSS *g'*, *r'*, *i'*, and *z'* (e.g., Fukugita et al. 1996). The conditions were not photometric and the field was too far south for the Pan-STARRS catalog, so absolute calibrations were performed by re-imaging the field on 2017 July 2 (when it was photometric) in the same four broadband filters. We corrected for the varying extinction using on-chip comparison stars, finding typical corrections of ~0.5 magnitudes, which is near the limit of what we have previously found to yield reliable lightcurves (Knight et al. 2012, Eisner et al. 2017).

---

[8]https://sungrazer.nrl.navy.mil/index.php?p=news/machholz_babies
[9]https://twitter.com/sungrazercomets/status/925042636460785665

TABLE 1
Summary of Comet 96P/Machholz 1 Observations and Geometric Parameters during our 2017/18 Observations[a]

| UT Date[b] | UT Range | ΔT (days)[c] | Tel.[d] | $r_H$ (au) | $\Delta$ (au) | PA (°)[e] | α (°)[f] | $\sigma_m$[g] (mag) | $\sigma_{cal}$[h] (mag) | Filters | Conditions |
|---|---|---|---|---|---|---|---|---|---|---|---|
| 2017 Apr 7 | 08:22-09:38 | -203.6 | SOAR | 3.179 | 3.393 | 240.7 | 17.1 | 0.021 | 0.07 | g′, r′, i′, z′ | cirrus |
| 2017 Jul 2 | 02:43-10:45 | -117.7 | SOAR | 2.283 | 1.537 | 289.0 | 21.3 | 0.004 | 0.02 | g′, r′, i′, z′ | photometric |
| 2017 Jul 3 | 08:38-09:11 | -116.6 | SOAR | 2.270 | 1.520 | 290.7 | 21.3 | 0.003 | 0.03 | r′ | photometric |
| 2018 Jun 25 | 00:02-09:19 | +240.2 | SOAR | 3.493 | 2.477 | 294.0 | 0.6 | 0.027 | 0.08 | g′, r′, i′, z′ | cirrus |
| 2018 Jul 10/11 | 23:21-04:32 | +256.1 | SOAR | 3.619 | 2.642 | 85.4 | 5.2 | 0.028 | 0.03 | r′, i′, z′ | cirrus |
| 2018 Jul 12 | 02:41-07:17 | +257.2 | NTT | 3.628 | 2.657 | 85.8 | 5.5 | 0.080 | 0.11 | r′ | clear |
| 2018 Jul 31/Aug 1 | 22:03-01:35 | +277.0 | WHT | 3.778 | 2.980 | 90.0 | 10.7 | 0.041 | 0.03 | r′ | clear |
| 2018 Aug 1/2 | 21:49-01:29 | +278.0 | WHT | 3.784 | 2.998 | 90.1 | 10.9 | 0.029 | 0.02 | r′ | photometric |
| 2018 Aug 2 | 21:47-22:51 | +279.0 | WHT | 3.792 | 3.017 | 90.3 | 11.1 | 0.073 | 0.09 | r′ | photometric |

*Notes.*
[a] All parameters were taken at the midpoint of each night's observations.
[b] UT date of the observations (note that some nights span multiple UT dates).
[c] Time since perihelion (2017 Oct. 27.96)
[d] Telescope: SOAR = Southern Astrophysical Research (4.2-m), NTT = New Technology Telescope (3.54-m), WHT = William Herschel Telescope (4.2-m).
[e] Position angle (PA) of the Sun.
[f] Phase angle.
[g] Average photometric (statistical) uncertainty in the r′-band magnitudes calculated for the night.
[h] Uncertainty in the absolute calibration of the r′-band magnitudes for the night.

### 2.1.2. 2017 July 2-3

Images taken on 2017 July 2 and 3 were obtained at SOAR with the SOAR Optical Imager (SOI)[2] which uses a mosaic of two e2v 2048 × 4096 pixel CCDs (Walker et al. 2003). On-chip 2 × 2 binning produced images with a pixel scale of 0.154 arcseconds pixel$^{-1}$. On July 2 we monitored the comet all night in r′ and occasionally took snapshots using the g′, i′, and z′ filters. A short sequence of r′ images was obtained on July 3. The conditions throughout both of these nights were photometric, allowing us to use SDSS standard stars to perform absolute calibrations (Smith et al. 2002).

### 2.1.3. 2018 June 25 and 2018 July 10/11

The observations on 2018 June 25 and July 10/11 made use of the Goodman Spectrograph Red camera at the SOAR telescope. On both nights, images were obtained using broadband g′, r′, i′, and z′ filters, however, due to technical problems the g′ filter images obtained on 2018 July 10/11 could not be used for further analysis.

Throughout our 2018 observations the comet was located in the midst of the Milky Way and, therefore, suffered from significant contamination from crowded star fields. Difference Image Analysis (DIA, Bramich 2008; Bramich et al. 2013) was implemented to recover useful data by removing background stars, as we previously used to extract useful data on 67P/Churyumov–Gerasimenko when it was located in crowded fields in 2014 (Snodgrass et al. 2016). The technique uses background templates (or reference images) of the fields, which were taken on 2018 July 11, and works by blurring them to match the seeing of each comet frame. The blurred reference image is then subtracted from the comet frames, removing the majority of the background stars as shown in *Figure 1*. The figure also demonstrates that saturated stars cannot be removed. Absolute calibration was carried out by measuring field stars in the reference image and comparing the observed magnitudes to those from the Pan-STARRS catalog PS1 Data Release 1 (Chambers et al. 2017). The PS1 filter system differs from SDSS by small amounts, and thus we corrected for this using the equations presented by Tonry et al. (2012).

We monitored the comet in r′ on both nights and took three sequences of g′, r′, i′, and z′ filters images on 2018 July 10/11 at times when the comet was located between field stars. As reference images were only taken in r′ the DIA method could not be applied to these data. We determined absolute calibrations on each image using the Pan-STARRS catalog.

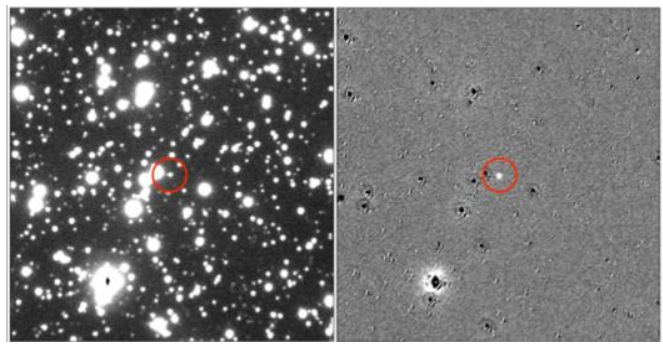

*Figure 1.* DIA image processing of the same initial image (left), taken on the night of 2018 July 10/11. The images show approximately 300,000 km across at the scale of the comet. The position of the comet is shown by the red circle. The star in lower left of each image cold not be removed properly due to it being saturated.

---

[2] SOI was used during this run to support the blue wavelength sensitivity needed by another program on these nights.

### 2.1.4. 2018 July 12

Snapshot $r'$-band images were obtained on 2018 July 12 using the 3.54-m New Technology Telescope (NTT) at La Silla, Chile, with the ESO Faint Object Spectrograph and Camera (EFOSC2) which employs a 2048 × 2048 pixel Loral/Lesser CCD (Buzzoni et al. 1984). On-chip 2 × 2 binning produced images with a pixel scale of 0.24 arcseconds pixel$^{-1}$. The Pan-STARRS catalog was used to perform absolute calibrations. Due to the conditions not being photometric, we monitored the brightness of seven field stars in order to correct for variations in transparency throughout the night (see methodology by Eisner et al. 2017 for details), yielding image-to-image corrections between -0.01 and 0.13 mag.

### 2.1.5. 2018 July 31 – August 2

Our final images were acquired using the 4.2-m William Herschel Telescope (WHT), in La Palma, Spain, on 2018 July 31 through August 2. These observations made use of the auxiliary port camera (ACAM; Benn et al. 2008), which employs an e2v 2k × 4k CCD. On chip binning of 1 × 1 yielded a pixel scale of 0.253 arcseconds pixel$^{-1}$.

On the first two of these nights, the comet observations suffered from significant contamination from background stars and thus DIA methods were applied in order to obtain coherent lightcurves. This was not necessary on 2018 August 2, due to the comet being fortuitously located in front of a Milky way dust lane, producing a field of view relatively free of stars. Field stars in the Pan-STARRS catalog were used to perform absolute calibrations on all three nights.

### 2.2. Comet measurements

Following the methodology of earlier papers (e.g., Knight et al. 2011, Eisner et al. 2017) we extracted the flux by centroiding on the comet nucleus and integrating inside a circular aperture. The aperture radii were independently chosen for each night to maximize the signal-to-noise ratio (S/N) whilst minimizing the contamination from field stars, but were typically around 1.5 times the typical FWHM. On nights where the DIA methods were applied to extract data (2018 June 24, July 10/11, July 31/August 1, August 1/2), the aperture sizes were automatically adjusted throughout the night in order to match the seeing, with an aperture radius set equal to the point spread function FWHM to maximize the S/N, and an aperture correction was applied (following the procedure applied by e.g., Snodgrass et al 2005, 2013). The median sky flux, calculated in an annulus with a radius significantly larger than the circular aperture, was subsequently subtracted from the measured flux of the comet. All observed $r'$-band magnitudes and their statistical uncertainties are tabulated and available as an online supplement. As will be discussed in the following section, we did not detect any coma so the absolute magnitude (apparent magnitude reduced to unit heliocentric and geocentric distances at zero solar phase angle), $H$, was calculated using standard asteroidal normalization (e.g., Jewitt 1991), assuming a linear phase correction of 0.04 magnitudes per degree.

## 3. NUCLEUS PROPERTIES

### 3.1. Assessment of Coma

In order to study properties of the nucleus, the extent of the contribution from the coma needed to be assessed. Individual images on all nights showed no evidence of a prominent coma or tail, and thus we created stacked images using a median of all the frames obtained on a single night. An example of this is shown in the left-hand panel of *Figure 2* for the night of 2017 July 1 with a total integration time of 20280 sec (5.6 hr). These stacked images also showed no signs of tail or coma.

Additionally, we compared the radial profiles of stacked images and single frame images of the comet to radial profiles of nearly untrailed field stars. All three were found to be consistent with one another in both our pre-and post-perihelion observations, further confirming that we were observing a bare nucleus. A representative example is shown in the right-hand panel of *Figure 2* for the night of 2017 July 1.

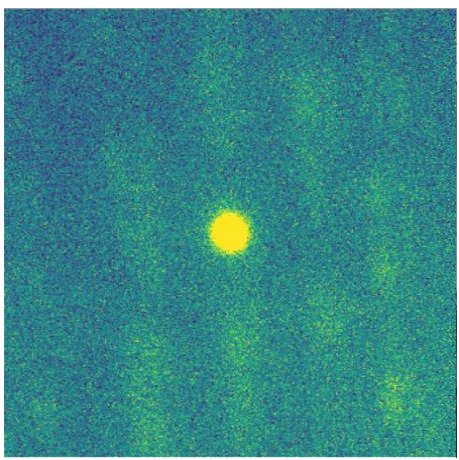
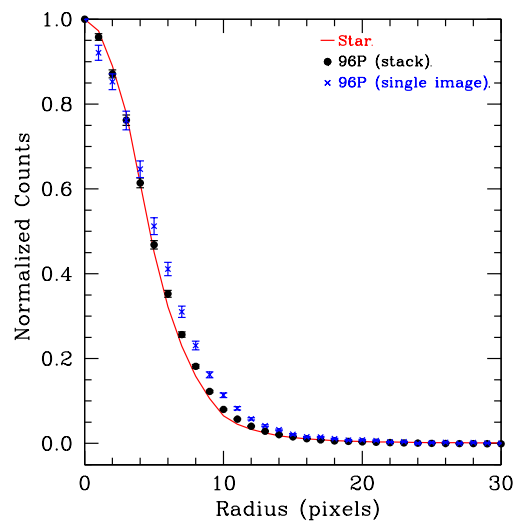

*Figure 2.* The left-hand panel shows a median stacked image from the night of 2017 July 1 with a physical size of 67,000 km at the comet and an equivalent exposure time of 20280 sec (5.6 hr). The stacked image is stretched to reveal faint structures, and shows faint stars trailing across the field of view. It highlights the lack of any signs of a coma or tail within the field of view. The right-hand panel shows the normalized radial profile also from 2017 July 1 of comet 96P in a single 120 sec exposure image (blue crosses), of the median stacked image with an equivalent exposure time of 20280 sec (black circles), and of an essentially untrailed field star in a 1 sec exposure (red line). The profile of the comet on the single image is slightly wider than the nightly stack due to variations in seeing.

### 3.2. Nucleus Size

Based on the assumption that we observed a bare nucleus we estimated its effective radius ($R_n$) from the absolute magnitude using the standard methodology devised by Russell (1916) and reformulated by Lamy (2004):

$$R_n = \frac{1.496 \times 10^{11}}{\sqrt{p}} 10^{0.2(m_\odot - H)} \qquad [1]$$

where $p$, $H$, and $m_\odot$ are the geometric albedo, absolute magnitude, and the apparent magnitude of the Sun, respectively and $R_n$ has units of meters. A value of -26.93 was used for the $r'$-band magnitude of the Sun, based on the color of the Sun ($V − R = 0.370$; Colina et al. 1996) and the Smith et al. (2002) $V$ to $r'$ transformation. Typical values for albedo (0.04) and nucleus phase angle correction (0.04 mag deg$^{-1}$; see *Section 3.4*) were assumed (e.g., Lamy et al. 2004). The average absolute magnitudes ($\bar{H}_{r'}$) and corresponding uncertainties are given in *Table 2*. For nights with both a peak and a trough, they were determined at the midpoint of the magnitude variation. On nights with incomplete/ambiguous lightcurves, the average absolute magnitude was estimated based on the rotational phasing (see Section 5). The average nucleus radius across all nights was found to be 3.41 ± 0.24 km, where the uncertainty is the standard deviation of the nightly results.

On nights where the broadband $r'$ data showed a clear peak and trough, the minimum axial ratio of the nucleus was determined based on the peak-to-trough maximum amplitude of the lightcurve. The peak-to-trough variation of the lightcurves and the corresponding axial ratios are tabulated in *Table 2*.

Both the average radius and axial ratio were compared to previously determined values. Licandro et al. (2000) observed a 'starlike' nucleus at $r_H$ = 4.89 AU in 1995 finding

TABLE 2
Summary of Comet 96P/Machholz 1's average nucleus properties

| Date[a] | $\bar{H}_{r'}$[b] | $R_n$ (km)[c] | Amplitude | Axis ratio[e] |
|---|---|---|---|---|
| 2017 Apr 7 | 14.5 ± 0.1 | 3.87 ± 0.2 | - | - |
| 2017 Jul 2 | 14.85 ± 0.05 | 3.30 ± 0.08 | 0.50 ± 0.05 | 1.58 ± 0.07 |
| 2017 Jul 3 | 14.85 ± 0.10 | 3.30 ± 0.15 | - | - |
| 2018 Jun 25 | 14.55 ± 0.1 | 3.78 ± 0.2 | - | - |
| 2018 Jul 10 | 14.85 ± 0.05 | 3.29 ± 0.08 | 0.30 ± 0.10 | 1.32 ± 0.12 |
| 2018 Jul 12 | 15.0 ± 0.1 | 3.1 ± 0.1 | - | - |
| 2018 Jul 31 | 14.80 ± 0.05 | 3.37 ± 0.08 | 0.35 ± 0.10 | 1.38 ± 0.13 |
| 2018 Aug 1 | 14.80 ± 0.05 | 3.37 ± 0.08 | 0.35 ± 0.10 | 1.38 ± 0.13 |
| 2018 Aug 2 | 14.80 ± 0.10 | 3.37 ± 0.15 | - | - |

*Notes.*
[a] UT date at the start of the observations.
[b] Estimated absolute magnitude at midpoint of the magnitude variation.
[c] Estimated nucleus radius.
[d] The maximum peak-to-trough amplitude for nights which yielded a clear peak and trough.
[e] Minimum axis ratio.

an equivalent radius of 3.6 km and axial ratio of 1.4. Meech (1996) reports an equivalent radius of 2.8 km and an axial ratio of 1.4 but provides no further information. Using these two reports, Lamy et al. (2004) note that an axial ratio of a/b ~ 1.5 could yield a spheroid with a = 4.3 km and b = 2.8 km, making both observations consistent with our $R_n$.

### 3.3. Nucleus Color

We obtained color measurements pre-, as well as post-perihelion on the nights of 2017 April 7, 2017 July 2, and 2018 July 10/11. On 2017 April 7 color observations, following the sequence $g'$, $r'$, $i'$, $z'$, $z'$, $i'$, $r'$, $g'$, were taken twice throughout the night. For each sequence, the mean magnitude per filter was used to calculate quasi-simultaneous colors $g'$-$r'$, $r'$-$i'$, and $i'$-$z'$. On 2017 July 1 $r'$-band observations were interleaved with observations in the $g'$, $i'$, and $z'$ -filters five times throughout the night. We interpolated the $r'$-band observations to the time of the respective $g'$, $i'$, and $z'$ observations to calculate colors, however, the resulting $g'$-$r'$, $r'$-$i'$, and $i'$-$z'$ are not quasi-simultaneous. As previously mentioned, we were unable to obtain useable images in the $g'$-filter on the night of 2018 July 10/11, and thus we were only able to calculate $r'$-$i'$, and $i'$-$z'$ colors post-perihelion.

These colors were converted to *B-V*, *V-R* and *R-I* colors using translations from Lupton (2005) to allow for direct

TABLE 3
Summary of Comet 96P/Machholz 1 color measurements

| UT date | Time | $g'$ | $r'$ | $i'$ | $z'$ | $g'$-$r'$ | $r'$-$i'$ | $i'$-$z'$ | *B-V*[a] | *V-R*[a] | *R-I*[a] |
|---|---|---|---|---|---|---|---|---|---|---|---|
| **2017 Apr 7** | 08:47 | 20.90 ± 0.069 | 20.47 ± 0.047 | 20.30 ± 0.054 | 20.23 ± 0.087 | 0.43 ± 0.083 | 0.17 ± 0.071 | 0.07 ± 0.102 | 0.62 ± 0.08 | 0.35 ± 0.08 | 0.39 ± 0.07 |
| | 09:24 | 20.75 ± 0.069 | 20.29 ± 0.042 | 20.14 ± 0.051 | 20.08 ± 0.074 | 0.46 ± 0.081 | 0.15 ± 0.066 | 0.07 ± 0.090 | 0.64 ± 0.08 | 0.37 ± 0.08 | 0.38 ± 0.07 |
| **Average**[b] | | | | | | 0.45 ± 0.02 | 0.16 ± 0.01 | 0.07 ± 0.01 | 0.63 ± 0.01 | 0.36 ± 0.01 | 0.39 ± 0.01 |
| **2017 Jul 1** | 03:54 | 19.10 ± 0.017 | 18.56 ± 0.016 | 18.41 ± 0.016 | 18.28 ± 0.019 | 0.54 ± 0.023 | 0.13 ± 0.023 | 0.11 ± 0.025 | 0.71 ± 0.03 | 0.42 ± 0.03 | 0.36 ± 0.02 |
| | 05:02 | 18.94 ± 0.016 | 18.42 ± 0.016 | 18.27 ± 0.016 | 18.32 ± 0.019 | 0.52 ± 0.022 | 0.18 ± 0.022 | -0.02 ± 0.025 | 0.70 ± 0.03 | 0.41 ± 0.03 | 0.41 ± 0.02 |
| | 06:36 | 18.90 ± 0.016 | 18.4 ± 0.015 | 18.17 ± 0.016 | 18.15 ± 0.017 | 0.50 ± 0.022 | 0.23 ± 0.022 | 0.03 ± 0.023 | 0.68 ± 0.02 | 0.40 ± 0.02 | 0.46 ± 0.02 |
| | 07:44 | 19.05 ± 0.016 | 18.54 ± 0.015 | 18.39 ± 0.016 | 18.31 ± 0.018 | 0.51 ± 0.022 | 0.17 ± 0.022 | 0.09 ± 0.024 | 0.69 ± 0.02 | 0.40 ± 0.02 | 0.40 ± 0.02 |
| | 09:03 | 18.88 ± 0.016 | 18.37 ± 0.015 | 18.23 ± 0.016 | 18.18 ± 0.017 | 0.51 ± 0.022 | 0.18 ± 0.022 | 0.09 ± 0.023 | 0.69 ± 0.02 | 0.40 ± 0.02 | 0.41 ± 0.02 |
| **Average**[b] | | | | | | 0.52 ± 0.02 | 0.18 ± 0.04 | 0.06 ± 0.06 | 0.69 ± 0.01 | 0.41 ± 0.01 | 0.41 ± 0.04 |

*Notes.*
[a] B-V, V-R, and R-I colors were calculated using translations from Lupton (2005).
[b] The mean color on a given night

comparisons with existing data sets. Photometric uncertainties were calculated from adding in quadrature the photon statistics (see column 9 of *Table 1*), uncertainties in the absolute calibration (see column 10 of *Table 1*), and uncertainties in the color translations (typically less than 0.01). All color measurements and their uncertainties are tabulated in *Table 3*.

Average colors were determined for each night in order to compare the color of 96P's nucleus to that of other known comet nuclei colors. The average colors are tabulated in *Table 3*, where the uncertainties are the standard deviations of all the individual colors measurements on a given night. It should be noted that the uncertainties in the individual measurements, which are also given in *Table 3*, are mostly substantially larger. The *V-R* color of comet 96P had previously been measured to be 0.3 ± 0.1 during the 1995/1996 apparition (Licandro et al. 2000) and 0.429 ± 0.027 (apparition unspecified; Meech et al. 2004). Both of these values are consistent with our *V-R* colors determined for the 2017/2018 apparition, within the uncertainties of individual measurements.

*Figure 3* shows a comparison of the average colors of 96P to those of the nuclei of Jupiter family comets and long period comets (Lamy et al. 2004, 2009; Lamy & Toth 2009; Hainaut et al. 2012, and references therein), as well as to the interstellar object 1I/'Oumuamua (Bannister et al. 2017; Jewitt et al. 2017) and comet 322P (Knight et al. 2016). The error bars represent the uncertainties calculated from the standard deviation of the measurements, which, as previously mentioned, are significantly smaller than the uncertainties in the individual color measurements. The colors of 96P for the two nights are consistent within the uncertainties of individual images. The *B-V* and *V-R* colors presented for 'Oumuamua were determined by Jewitt et al. (2017). As they did not report on a *R-I* color, we calculated this value based on data of Bannister et al. (2017), which resulted in significantly larger error bars, as can be seen in *Figure 3*. Bolin et al. (2018) also reported colors of 'Oumuamua, however, as their uncertainties span the plotted region, we have elected not to show these. The plot shows a notable difference between the color of comet 96P and the average color of the nuclei of Jupiter family comets, exhibiting a generally bluer color. Intriguingly, the plot of B-V vs. V-R (top panel of *Figure 3*) suggests some similarities between the colors of comet 96P, 1I/'Oumuamua, and 322P, which we will return to in *Section 6*. The color of 96P is also similar to that of long-period comets (Halley-type comets) in all three color metrics, which will also be discussed in *Section 6*. The results show no evidence for rotational variation of the colors or for pre-/post-perihelion changes.

### 3.4. Nucleus Phase Function

As our observations spanned a diagnostic range of phase angles from 0.6° to 21°, we attempted to determine the nucleus phase function. We applied two techniques: a simple approach using the estimated midpoint of the magnitude variation each night (as discussed in *Section 3.2*), and a more sophisticated, stochastic approach that uses all data points and their associated errors (see Kokotanekova et al. 2017, 2018). While the methods yielded results that were generally consistent with each other, the derived phase function slope was highly sensitive to precise absolute calibrations. Due to the overall challenging observations (weather, dense background fields, moon) which necessitated a variety of reduction techniques to process the different nights, we concluded that the absolute calibrations are not known well enough to unambiguously determine the phase function and have elected to use 0.04 mag deg$^{-1}$ throughout this paper. However, there is some evidence that a flatter phase function (0.02-0.03 mag deg$^{-1}$) might be valid, which would be consistent with the finding of Kokotanekova et al. (2018) of a shallow phase dependences for larger and less active nuclei. There is also some indication of an opposition surge based on the 2018 June 25 data when 96P was close to minimum phase angle (0.6°). While this night was chosen specifically to allow this test, the very dense stellar background in the Milky Way and the strong scattered light from the proximity to a nearly full moon resulted in the largest uncertainties of our dataset, and we are thus reluctant to draw a firm conclusion from this night.

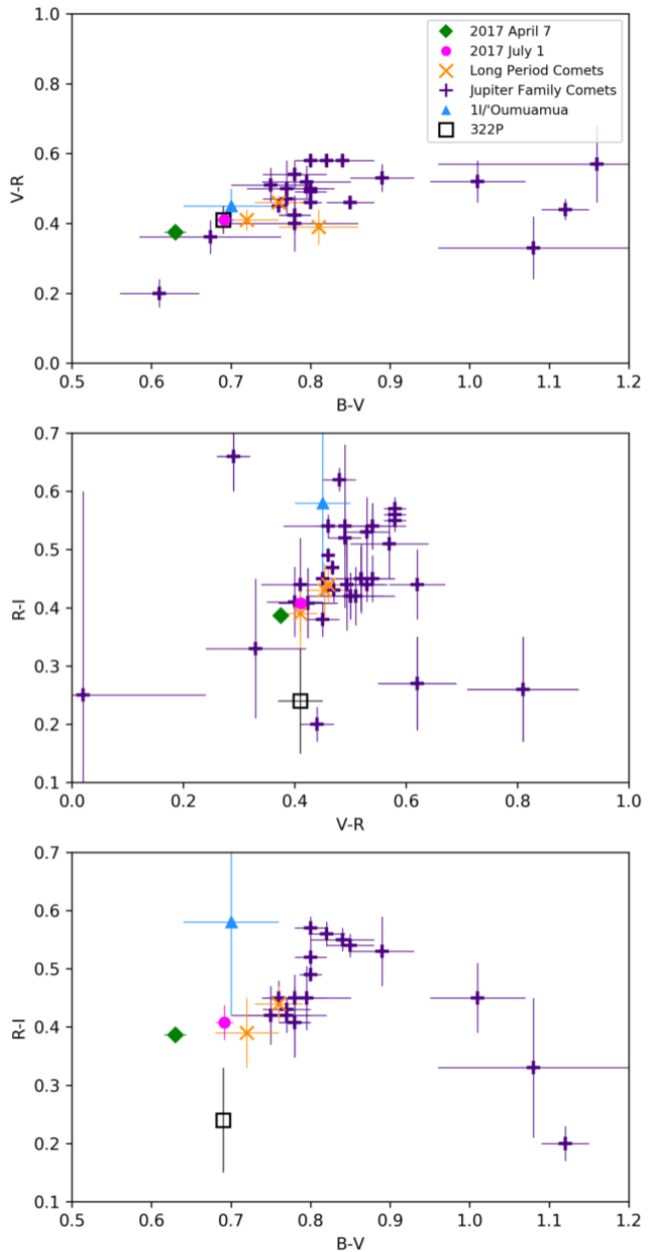

*Figure 3.* Average colors for 96P (pink circle and green diamond) compared to colors of long period comets (yellow crosses) and Jupiter family comets (indigo plus signs), 1I/'Oumuamua (light blue triangle) and 322P (black squares). See text for references. Our SDSS colors for 96P were translated to Johnson colors following Lupton (2005) for comparison with the other data sets. Some of the comets do not have all three color measurements and therefore only show up in one out of the three plots.

## 4. SEARCH FOR FAINT COMPANIONS

As discussed in the introduction, 96P has been linked to two groups of near-Sun comets and is believed to have had frequent fragmentation, with *SOHO* images of 96P at perihelion revealing faint accompanying fragments in 2012 (two fragments) and 2017 (three fragments). The fragments seen in 2017 were on orbits several hours ahead of comet 96P and, therefore, would be expected to be well outside of our fields of view by the time of our ground based observations in 2018. Nonetheless, if the fragmentation seen by *SOHO* in

2017 occurred relatively recently, the fragments could have still been within our field of view during our 2017 ground based observations. It is also possible that there were fainter or other recently produced fragments in our fields of view.

To search for possible companions, we created deep images by median-combining all images on a given night, effectively removing all stars. The best nights were 2017 July 1 (pre-perihelion) and 2018 July 10/11 (post-perihelion) since they combined a long time on the comet with good seeing, and minimal scattered light from the moon. Stacked images on 2017 July 1 totaled 20280 sec (5.6 hr) over a 10 × 10 arcsec region centered on the comet (left-hand panel of *Figure 2*) and 15360 sec (4.3 hr) over a 30 × 30 arcsec region, while our 2018 July 10/11 stacked image totaled 15540 sec (4.3 hr) over a 300 × 300 arcsec region centered on the comet. There is a steep decrease in effective exposure time as the comet-centric distance increases for our 2017 data because we conducted large dithers and have restricted the stacked images to only the CCD on which 96P was observed in case there are small differences in detector sensitivity or calibrations. The 2018 data did not have such problems and consequently go equally deep at all distances.

No faint companions were seen in any of the deep images. We estimated limiting magnitudes in these deep images by scaling the exposure time and signal-to-noise relative to our individual 96P exposures on each night and setting the detection threshold at a signal-to-noise of 10. The 2017 July 1 stacked images rule out any accompanying fragments down to a magnitude of $m_{r'} \sim 26$, equivalent to a radius of ~100 m for the same assumptions as the nucleus within the 30 × 30 arcsec region. The 2018 data are severely limited by sky uncertainty, and the 2018 July 10/11 stacked image only rules out accompanying fragments down to a magnitude of $m_{r'} \sim 21$, corresponding to roughly 2 km radius objects. As this is nearly as large as 96P itself, searches for newly created accompanying fragments in 2018 are very unconstraining given that estimates of the fragments seen by *SOHO* are of the order of a few 10s of meters in diameter (Battams & Knight 2017).

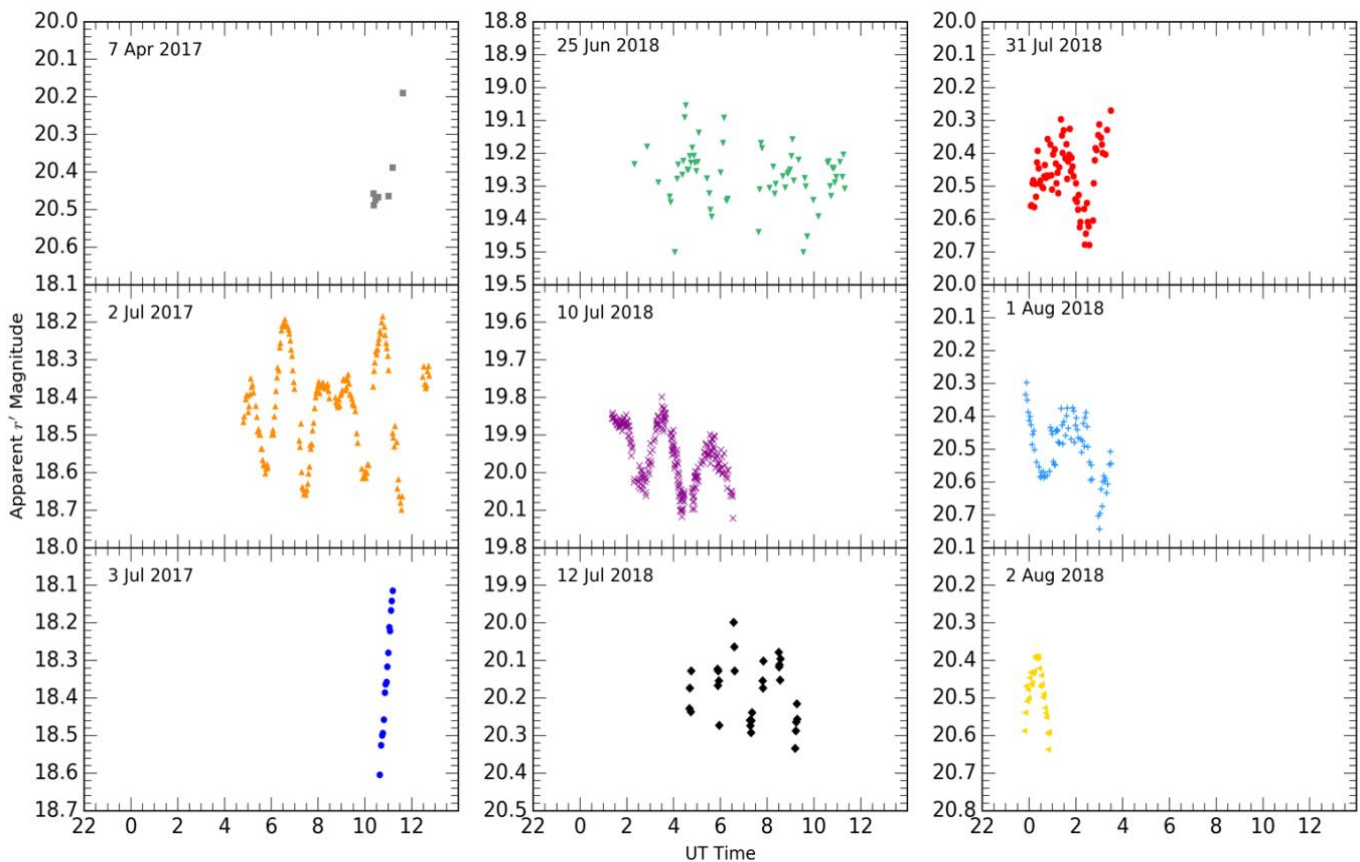

***Figure 4.*** Apparent *r'*-band magnitudes ($m_{r'}$) plotted as a function of UT on each night. The magnitudes have had absolute calibration, extinction correction and field star corrections applied as described in sections 2.1.1 to 2.1.5. The dates represent the UT date at the time of the first observation. Uncertainties are not shown on these and subsequent lightcurve plots as they were typically smaller than the data points; however, the average photometric and calibration uncertainties for each night are given in *Table 1* and the uncertainties of individual measurements are given in the supplementary table.

## 5. LIGHTCURVE ANALYSIS

### 5.1. Lightcurve Shape

As shown in *Figure 4*, the lightcurve is double-peaked with one peak pronounced and the other broad and flat, and one trough significantly shallower and 'rounder' than the other. This asymmetry is likely due to the nucleus shape deviating from that of a simple triaxial ellipsoid, for example it may have sharp edges, large boulders or flat planes (e.g., Lowry et al. 2012). The asymmetric shape of the lightcurves improved our ability to phase the data as it allowed us to eliminate half-cycle aliases (see *Section 5.3*).

The pre-perihelion lightcurve is much better defined than the post-perihelion lightcurve due to the significantly higher S/N and less crowded fields. While the general pre- and post- perihelion shape is similar, there are several distinct differences, such as that the flat region is shorter in duration and the amplitude difference between the two peaks is smaller in the post-perihelion data. A likely explanation for these pre- and post perihelion differences is that we are viewing the comet from a different angle (maximum difference of 60°), and thus see a noticeable change in the shape of the lightcurve. Additionally, the peak-to-trough amplitude decreases from 0.50 ± 0.07 mag to 0.30 ± 0.1 mag between July 2017 and July 2018 (*Table 2*). Due to the differences in viewing geometry between the two epochs of observations, the change in amplitudes suggest that we are viewing the nucleus at a different aspect ratio. However, additional observations at other viewing geometries would be needed to significantly constrain the pole orientation.

### 5.2. Pre-Perihelion Lightcurve (2017)

In order to determine the rotation period of comet 96P, we looked for periodic variations in the lightcurves. As we do not have knowledge of the pole orientation we only considered the synodic rotation period. However, for most pole orientations, the synodic period should have been within ~2 sec of the sidereal period given that the phase angle bisector angle (Harris et al. 1984) changed by only ~0.3° between 2017 July 2 and 3. The observations from 2017 July 2 spanned ~8 hours (see *Figure 4*) and exhibited a double-peaked lightcurve that repeated after ~4.1 hours. This rotation period was further constrained by superimposing the partial lightcurve from the subsequent night, with the data phased to a 'trial' period and zero phase set at perihelion (2017 October 27.96). Systematic iterations through the trial period allowed us to make direct 'better-or-worse' comparisons between different rotation periods, as shown in *Figure 5*. Using this method, we were able to constrain the pre-perihelion rotation period to 4.10 ± 0.03 hour. As the data taken on 2017 July 3 (navy blue circles) do not show signs of a trough nor a peak, the uncertainty in the rotation period is dominated by the uncertainty in the absolute calibration. These uncertainties affected the solution by ~0.02 hours, whilst the uncertainty in the timing for any given absolute calibration was of the order of ~0.01 hours. Attempts were also made to constrain the rotation period using the observation from 2017 April 7, however these data were too sparse and the time interval was too long to yield useful constraints, since numerous tightly spaced aliases were possible.

We independently searched for the period using phase dispersion minimization (PDM; Stellingwerf 1978). PDM is a widely used numerical technique for analyzing sparse, non-sinusoidal lightcurves as it does not require uniformly sampled data. The 'true' period is indicated by a local minimum in the phase dispersion minimization plot ($\theta$), as shown in *Figure 6* at 4.097 ± 0.015 hours for the combined 2017 July 2 and 3 data. The uncertainty was quantified by determining the range in the minima of the PDM when the absolute magnitudes on both nights were shifted by their uncertainties in the absolute calibration. This rotation period is in excellent agreement with the rotation period obtained by inspection of the phased lightcurves. We conservatively use the less restrictve period (from phasing) for this epoch.

### 5.3. Post-Perihelion Lightcurve (2018)

In the same vein as for the pre-perihelion lightcurves we phased the post-perihelion data and took an iterative approach to scan through a large number of trial periods. This was done using data from the nights of 2018 July 10/11, July 12, July 31/ August 1, August 1/2, and August 2 as shown in *Figure 7*. Due to low signal-to-noise, the data from 2018 June 24 were not found to be constraining and were thus not used for phasing. We found a best period of 4.096 ± 0.002 hr through visual phasing of the data.

In order to aid with the comparison and to yield a more robust PDM result, we applied small shifts to the magnitudes of all the lightcurves in order to bring their extrema into approximate alignment with the lightcurve from 2018 Jul 10/11, which we found to be the most constraining night. In order to align the data from 2018 July 12, we applied an offset of -0.15. Whilst this shift is large, it is consistent with the uncertainty in the absolute calibration. The lightcurves from 2018 July 31/August 1, August 1/2, and August 2 were all offset by a magnitude of +0.08 magnitudes.

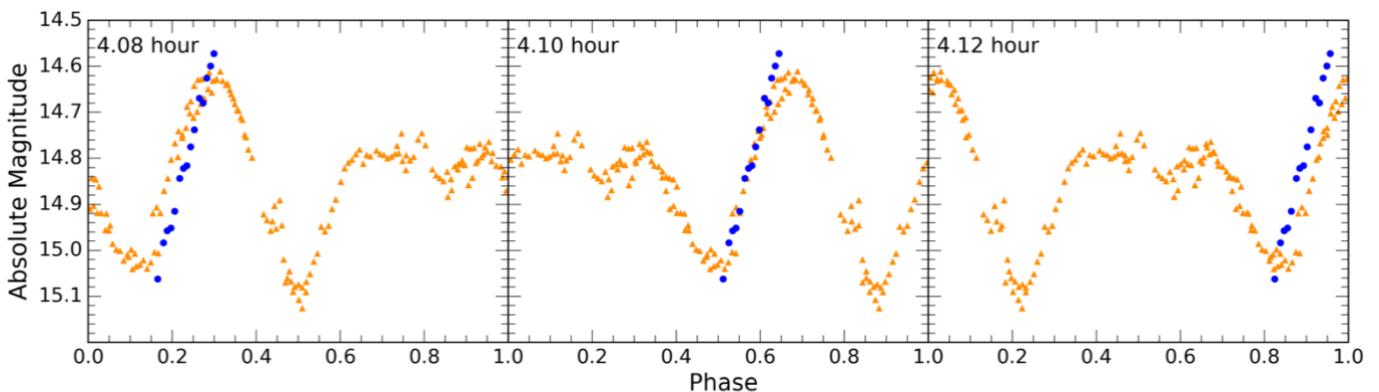

*Figure 5.* Our reduced pre-perihelion data ($H_{r'}$) for 2017 July 2 (orange circles) and July 3 (navy blue triangles) phased to 4.08 hr (left), 4.10 hr (middle), and 4.12 hr (right). By iterating though the different periods, we found a best period of 4.10 ± 0.03 hr, where the uncertainty is dominated by the uncertainties in the absolute calibration (± 0.02 magnitudes). Zero phase was set at the time of perihelion (2017 October 27.96).

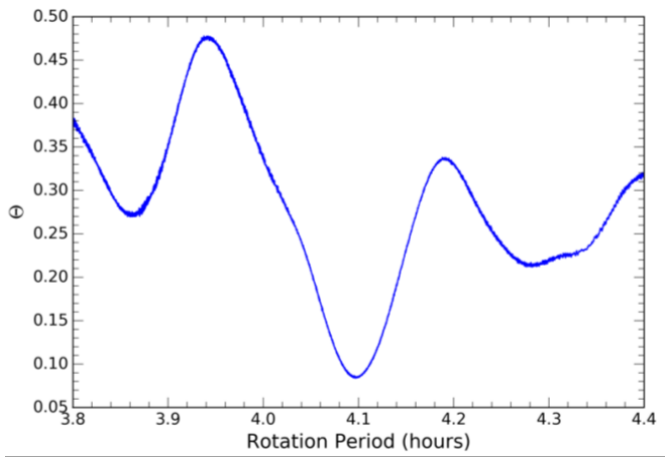

*Figure 6.* Phase dispersion minimization diagram for the pre-perihelion 2017 July 2 and 3 data. The data curve yields a clear minimum at a period of 4.097 ± 0.015 hours.

The need for these shifts is likely due to the differences in calibration techniques used on different nights, notably the differences in calibration of the DIA processed images and the traditional photometry images. Further differences could have resulted from using different aperture sizes, which were chosen based on the seeing on a given night. The effect of aperture size on the amplitude can be seen in the August 1/2 lightcurve (light blue plus signs in *Figure 7*), where the peaks and troughs are not clearly defined despite the maximum and minimum amplitudes being more extreme than on the other nights.

In addition to the misalignment of the peaks and troughs, there is a clear misalignment of the magnitudes at the beginning and end of the night on 2018 Jul 10 (purple crosses), located at a phase of ~0.4 in the left panel (*Figure 7*). Whilst using larger aperture sizes can correct for this misalignment, it also results in significant contamination from field stars, and we have thus elected to not correct for this effect.

The PDM plot for the post perihelion data (*Figure 8*) displays a minimum centered on 4.095 ± 0.011 hours. The local minima immediately surrounding 4.095 hours were rejected because some nights were clearly a half-cycle out of phase, whilst the next local minima moving away from the overall smallest value of θ clearly showed a misalignment of the data obtained on 2018 July 10/11 and on August 1/2. The alignment of the phased data became significantly worse as the rotation period continued to deviate from 4.096 hours, thus we are able to eliminate these aliases. The uncertainty in the PDM was assessed using Monte Carlo error analysis in a similar manner to that described in Kokotanekova et al. (2017). This method varies the magnitude of each data point by a random number from a Gaussian distribution with a standard deviation determined by the uncertainty in each point and a mean of zero. This was run 10,000 times and the standard deviation of all the minima taken to be the overall uncertainty.

The post-perihelion data obtained in 2018 June - August exhibited significantly lower signal-to-noise than the 2017 data due to the worse geometry, i.e. the heliocentric and geocentric distances were larger, during these later observations. Furthermore, the comet was situated in the midst of the Milky Way and was located within ~90° degrees from a ~70% illuminated moon on 25 June 2018. Thus, our observations suffered from significant contamination from a crowded field and scattered light. Even though the scatter in the post-perihelion lightcurves was greater than that of the pre-perihelion data, the larger time baseline allowed us to determine the rotation period to a higher degree of accuracy, yielding a value of 4.096 ± 0.002 hr. Furthermore, despite the noise, we were able to distinguish half-cycle differences, allowing us to avoid aliases. As in 2017, the viewing geometry changed very little between our observations, with the phase angle bisector differing by only ~3° from 2018 Jun 25 to 2018 Aug 2, and yielding likely synodic and sidereal periods in agreement to within ~1 sec.

6. DISCUSSION

Our observations throughout the 2017/2018 apparition suggest little or no change in the rotation period during this passage, with a change of <0.036 hours (130 sec, or 2.2 min). At first glance, this lack of change is not surprising based on the large size of the nucleus, 3.41 ± 0.24 km, and the low active fraction of the comet (e.g., Combi et al. 2011). Similar results have been found for other large, relatively inactive nuclei, e.g., 10P/Tempel 2 (Knight et al. 2011), 49P/Arend-Rigaux (Eisner et al. 2017), and 14P/Wolf, 143P/Kowal-Mrkos, and 162P/Siding Spring (Kokotanekova et al. 2018).

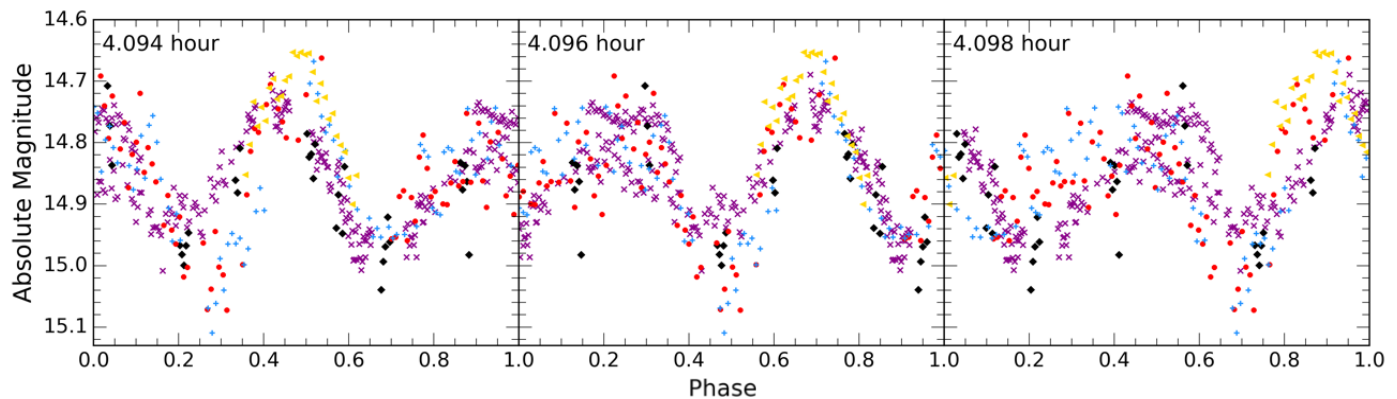

*Figure 7.* Our reduced pre-perihelion data ($H_{r'}$) for the nights of 2018 July 10/11, July 12, July 31/August 1, August 1/2, and August 2 phased to 4.094 hr (left), 4.096 hr (middle), and 4.098 hr (right). Colors and symbols are as given in *Figure 1*. Zero phase was set at the time of perihelion 2017 Oct. 27.96. Data obtained on 2018 July 12 and 2018 July 31 through August 2 have been offset by -0.15 and +0.08, respectively.

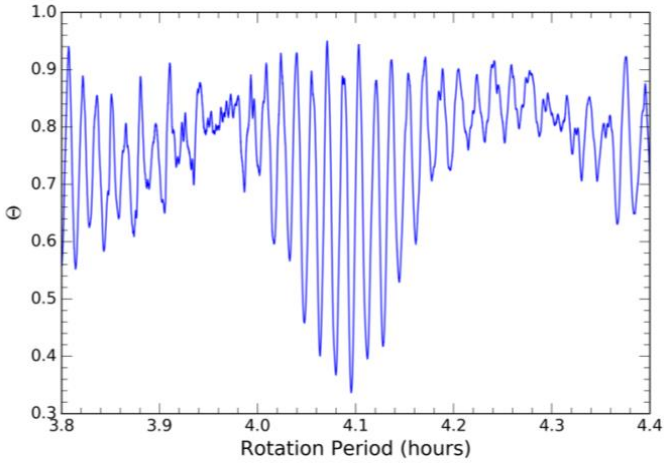

*Figure 8.* Phase Dispersion Minimization diagram for the post-perihelion data shown in *Figure 7* highlighting a minimum centred on ~4.095 ± 0.011 hours.

However, 96P is very active near perihelion, with $Q_{H2O}$ ~4 × $10^{29}$ molecules $s^{-1}$ (Combi et al. 2019). We would expect this activity to exert a significant torque on the nucleus and thus potentially result in a measurable change in rotation period. Furthermore, Mueller & Samarasinha (2018) argued that 96P is likely to exhibit a large change in rotation period based on models that relate changes in rotation period of Jupiter family comets to other physical parameters. They predict the maximum change in rotation period of comet 96P to be 9.1 minutes per apparition based on previous estimates for nucleus size and rotation period. Updating their calculation with our new rotation period and nuclear radius lowers this to 2.5 min, although for their less extreme torque cases, the predicted change in rotation period is less than our upper limit.

In order to examine this further, we compared the rotation period of 96P during the 2017 apparition to previous results found in the literature. To the best of our knowledge the only other time that attempts were made to quantify the rotation period was during the 1995/1996 apparition, where Meech (1996) reported a period of 6.38 hours. This is notably longer than the periods determined for the most recent apparition (4.10 ± 0.03 hours and 4.096 ± 0.002 hours before and after perihelion, respectively), possibly suggesting a significant change in rotation period over the course of the 21 years between the two measurements. If we assume that the rotation period decreased by 9.1 minutes with every apparition since 1996, as predicted by Mueller & Samarasinha (2018), we obtain a total change of 36.4 minutes prior to the 2017 perihelion passage, bringing the period to ~5.77 hours. This value remains significantly different from our determined value for the 2017 apparition. As Meech's 1996 data are not shown, one possible explanation is that an alias near 4.1 hour exists. Notably, if their period was a 3:2 alias, it would imply a period of ~4.25 hours. To get to our period of 4.1 hours would require a much more reasonable change of ~2.2 min per orbit, which is within our uncertainty and therefore not necessarily detectable.

Alternatively, and less likely in our opinion, the rotation period of comet 96P could have drastically changed due to processes that are not considered by the model presented by Mueller & Samarasinha (2018). Certain processes, for example, could be more prominent due to the unusually small perihelion distance that comet 96P reaches. At perihelion, equilibrium temperatures reach ~800 K and subsolar temperatures can exceed 1000 K, potentially resulting in the sublimation of refractory materials (e.g., Mann et al. 2004).

A further possibility is that the rotation period has changed as a result of fragmentation. Comet 96P is known to fragment, with dynamical evidence that it is the parent of at least two groups of near-Sun comets (Ohtsuka et al. 2003, Sekanina & Chodas 2005). Additionally, two and three fragments were observed with *SOHO* during the 2012 and 2017 apparitions, respectively. The process of fragmentation can cause changes to the moment of inertia of the nucleus, possibly leading to a change in the rotation period (e.g., Samarasinha et al. 2004). Even though fragmentation may result in a change in rotation period, the effect is likely small in the case of comet 96P since the fragments are apparently dwarfed by the main body (based on their apparent brightnesses in SOHO images). Nonetheless, such an event would also lead to more exposed surface ices. Vigorous activity from these newly exposed ices might result in a torque that acts to change the rotation period far more efficiently than the fragmentation itself. We think that this latter explanation is unlikely as freshly exposed ice should result in increased gas production, which was not seen by Combi et al. (2019), who report similar *SOHO-SWAN* water production rates for the 1996-2012 apparitions.

Water production rates were further used to investigate the lack of an observable coma. In *Section 2.3* we showed that the contribution from the coma is negligible and thus that we can assume that we are observing a bare nucleus. This is somewhat surprising based on extrapolation of the water production rates determined by Combi et al. (2011). They found that the water production rate varies as 2.5 x $10^{27}$ $r_H^{-2.5}$ before perihelion and as 4.6 x $10^{27}$ $r_H^{-1.9}$ after perihelion in units of molecules per second. Assuming that this relationship holds beyond the distances observed by *SOHO-SWAN* (out to $r_H$ ~ 0.8 AU), and using Schleicher (2008)'s dust-to-gas measurements, we would expect the coma flux to be at least as bright as the nucleus flux during our smallest $r_H$ observations at 2.270 AU on 2017 July 2. The "total" magnitudes reported to the Minor Planet Center (MPC) for 96P are consistent with our findings that 96P had minimal coma. An unweighted fit to all total magnitudes in the MPC database at $r_H$ < 2 AU, normalized to $r_H=\Delta=$ 1 AU and with a linear phase angle correction of 0.04 mag per degree, suggests the brightness varies proportional to $r_H^{-4.5}$, and the "total" magnitudes converges with the "nucleus" magnitudes around 2 AU. Although sparse, reported magnitudes at larger heliocentric distances are suggestive of no coma. We caution over-interpreting such aggregated MPC data due to the variety of techniques, filters, and aperture sizes used; however, they support our conclusion that 96P is essentially inactive by 2.3 AU. The lack of detectable coma could be due to a strong seasonal effect i.e. the active regions are located as such that they turn on/off rapidly somewhere between 0.8 and 2.3 AU, or due to the comet possessing an even lower dust-to-gas ratio at 2.3 AU than found by Schleicher (2008) at 1.8 AU (which was already low relative to other comets in the A'Hearn et al. 1995 database); for example, we do not see dust but there is still gas that we do not detect.

Based on the assumption that we are observing a bare nucleus, we investigated the color of 96P during the most recent apparition. Whilst the color itself does not reveal much in terms of the physical properties of the nucleus, the direct comparison of this value to that of other bodies in the solar system can disclose information on the origin and/or evolution of the comet. Our results revealed that the nucleus of comet 96P is significantly more blue than the average nuclei of Jupiter family comets. This could, once again, be a result of the intense heat that the comet endures during its close approach to the Sun, and could be indicative of a more evolved comet surface. This argument is strengthened by the similarity in color to 322P, which has a $q = 0.05$ AU but whose cometary origin in uncertain (Knight et al. 2016).

The color of 96P is also notably similar to that of the two Halley-type comets that have published nuclear colors, comet 1P/Halley and C/2001 OG108. Although it is dangerous to draw conclusions from a sample size of two comets, at face-value the likenesses in all three color metrics could indicate that 96P is a Halley-type comet (recall that 96P's peculiar resonance with Jupiter precludes making a firm determination of its origin) and that this color is a signature of Halley-type comets or Oort cloud objects. Alternatively, the color could be an evolutionary trait, with comets becoming less red over time. Compared to Kuiper Belt Objects/Centaurs, comets appear to lose more ultra-red matter when they become active, possibly due to resurfacing (Jewitt 2018). This, however, may not apply to comets that reach extremely close perihelion distances due to less fallback of material.

Comet 96P's nuclear properties were further compared to those of other comets. As already discussed, the nucleus of comet 96P rotates with a period of ~4.1 hours, placing it amongst the fastest rotating comet nuclei known (Kokotanekova et al. 2017). Combined with the axial ratio of at least ~1.5:1, this suggests that either its density is unusually high (e.g., >0.6 g cm$^{-3}$ for a strengthless body) or that it has more internal strength than is typically attributed to comet nuclei. The only unquestionably cometary object thought to have a significantly more extreme combination of rotation period and axial ratio is Component C of 73P/Schwassmann-Wachmann 3, whose rotation period was measured after it underwent rampant fragmentation (Drahus et al. 2010). However, other observations (Nolan et al. 2006, Dykhuis et al. 2012) report rotation periods for 73P-C of at least 10 hr, and 96P may, in fact, be the most extreme rotator known. Thus, it seems likely that 96P may currently be at, or near, its rotational spin limit. We speculate that rotationally driven fragmentation may be a common, or even dominant, source of the dynamically related objects which appear to have split from 96P both recently (e.g., the fragments seen by SOHO in 2012 and 2017) and centuries ago (e.g., the "Machholz Complex", cf. Ohtsuka et al. 2003, Sekanina & Chodas 2005).

Two other potentially cometary objects stand out as having extreme combinations of rotation period and axial ratio: comet 322P/SOHO 1 (Knight et al. 2016) and interstellar object 1I/'Oumuamua (e.g., Meech et al. 2017, Drahus et al. 2018, Fraser et al. 2018). Intriguingly, both objects have similar $V-R$ and $B-V$ colors to 96P. Although 'Oumuamua was initially thought to be asteroidal due to the lack of detectable coma (e.g., Jewitt et al. 2017, Knight et al. 2017, Meech et al. 2017, Bolin et al. 2018), later astrometric analysis by Micheli et al. (2018) revealed unequivocal non-gravitational acceleration which the authors attributed to cometary outgassing. Micheli et al. concluded that such activity could have gone undetected if 'Oumuamua was depleted in CN relative to water by a factor of at least 15 and had large dust grains not detectable at visible wavelengths. As previously noted, 96P was suggested to have potentially been of interstellar origin due to its severe CN depletion relative to water (by a factor of 72; Schleicher 2008) and it has a low dust to gas ratio as derived from visible observations. Thus, the additional similarities between the nucleus of 96P and 'Oumuamua might strengthen the argument for 96P having had an interstellar origin (and that these surface properties were acquired during millions of years of cosmic-ray bombardment between systems, rather than having been due to evolution in our solar system, e.g., Fitzsimmons et al. 2018). Another possible interpretation is that, as speculated by Ye et al. (2017) and Ćuk et al. (2018), 'Oumuamua may have been ejected from its natal system during a close passage to its host star (or stars if a binary). In that scenario, the properties observed as it passed through our solar system could have been caused by evolutionary effects due to repeated close approaches to its original parent star, suggesting that 96P's unusual properties are evolutionary rather than indicative of the environment in which it formed. As exciting as such speculation is, the similarities between 96P and 'Oumuamua could simply be coincidental, since the $R-I$ colors, though poorly constrained, are quite different and may be diagnostic of compositional differences due to the presence of silicate absorption at ~1 micron (e.g., Reddy et al. 2015 and references therein)

As noted in the introduction, 322P was a motivator of the current investigation due to its combination of very small perihelion distance (0.054 AU), unusual colors, and fast rotation period. The similarity of 96P to 322P (and as just discussed, possibly 'Oumuamua) in all three aspects suggests that there may be a trend with perihelion distance, perhaps due to surface properties changing as a result of the intense heating near perihelion. In particular, annealing of the upper layers may give a comet strength, allowing it to sustain a spin rate above the nominal rotational breakup limit. Knight et al. (2016) showed that 322P had a high albedo (0.09-0.42); if this, too, is a result of the small perihelion distance, then 96P may be substantially smaller than has otherwise been assumed. A smaller size would also mean its active fraction is substantially larger than estimated by Combi et al. (2011), implying it is much less evolved. Further investigation of 96P's nucleus properties in order to investigate evolutionary trends with perihelion distance would be highly desirable since 322P and 96P are the only periodic comets observed by traditional (i.e., non-coronagraphic) telescopes with perihelion distances smaller than 2P/Encke's 0.336 AU.

## 7. CONCLUSION

We observed comet 96P/Machholz 1 on a total of 9 nights both pre- and post-perihelion during its 2017/2018 apparition. Throughout our observations it did not exhibit any evidence of a coma, a somewhat surprising result given that we observed at heliocentric distances as low as 2.3 AU, and suggesting that 96P has either strong seasonal activity or that its upper layers are depleted in volatiles due to repeated close perihelion passages. We did not see any evidence of the fragments seen accompanying 96P in the *SOHO* fields of view around perihelion in 2012 and 2017. Our best upper limit for such fragments is a radius of ~100 m, though the relatively small field of view likely does not meaningfully constrain their time of formation.

We obtained complete rotational lightcurves both pre- and post-perihelion, finding a double-peaked lightcurve with one strong peak and one wide, flat peak, and two clean minima. We measured rotation periods of 4.10 ± 0.03 hours and 4.096 ± 0.002 hours before and after perihelion, respectively. Thus, to within the uncertainties, the rotation period did not change during the apparition. Based on the assumption that we observed the bare nucleus, we estimated the effective radius to be 3.41 ± 0.24 km, and the minimal axial ratio to be 1.6±0.1:1. This is the third fastest rotation period ever measured for a comet nucleus and, along with the axial ratio, suggests that either its density is unusually high or that 96P has more internal strength than typical comet nuclei. The comet may be at or near its rotational spin limit, suggesting a possible mechanism for the shedding of the small fragments seen near perihelion.

We also measured nucleus colors both pre- and post-perihelion, finding average values of $g'-r' = 0.50 ± 0.04$, $r'-i' = 0.17 ± 0.03$, and $i'-z' = 0.06 ± 0.04$. These colors are bluer than typical Jupiter family comet nuclei, but similar to the two published Halley-type comet nuclei, potentially indicating that 96P evolved from a Halley-type orbit. If such a link could be firmly established, it might indicate that blue nucleus colors are an intrinsic characteristic of such orbits. Similar blue colors have also been observed for near-Sun comet 322P/SOHO and for interstellar object 1I/'Oumuamua, suggesting a possible evolutionary cause due to either a small perihelion distance or (less likely) space weathering during an interstellar journey.

The high inclination of 96P causes it to be very far south during most of its orbit, limiting the telescopes with which it can be effectively studied. Its next apparition in 2023 will be especially poor, but the 2028 apparition will be its best since discovery, when it passes just 0.32 AU from Earth prior to perihelion. We encourage further investigations of this beguiling object, and especially advocate the use of *James Webb Space Telescope* for thermal infrared studies to constrain the size, albedo, and grain size distribution.

**Acknowledgments**

We thank the referee for a thorough and helpful review and Carlos Corco, Patricio Ugarte, Jacqueline Seron, Juan Espinoza, Alfredo Zenteno, Regis Cartier, Cesar Briceno, and Sean Points for help obtaining our SOAR observations. N.E., M.M.K., and M.S.P.K. were supported by NASA Near Earth Object Observations grant No. NNX17AK15G. This research made use of Astropy, a community-developed core Python package for Astronomy (Astropy Collaboration, 2013), as well as PyAstronomy. This work also used SAOIMAGE DS9, developed by Smithsonian Astrophysical Observatory and the 'Aladin sky atlas' developed at Centre de Données astronomiques de Strasbourg (CDS), Strasbourg Observatory, France (Bonnarel et al. 2000). Additionally, GMOS standard star finding charts prepared by the AAO ITSO Office in conjunction with Macquarie University's PACE (Professional and Community Engagement) program were used.

The Pan-STARRS1 Surveys (PS1) and the PS1 public science archive have been made possible through contributions by the Institute for Astronomy, the University of Hawaii, the Pan-STARRS Project Office, the Max-Planck Society and its participating institutes, the Max Planck Institute for Astronomy, Heidelberg and the Max Planck Institute for Extraterrestrial Physics, Garching, The Johns Hopkins University, Durham University, the University of Edinburgh, the Queen's University Belfast, the Harvard-Smithsonian Center for Astrophysics, the Las Cumbres Observatory Global Telescope Network Incorporated, the National Central University of Taiwan, the Space Telescope Science Institute, the National Aeronautics and Space Administration under grant No. NNX08AR22G issued through the Planetary Science Division of the NASA Science Mission Directorate, the National Science Foundation grant No. AST-1238877, the University of Maryland, Eotvos Lorand University (ELTE), the Los Alamos National Laboratory, and the Gordon and Betty Moore Foundation.

*Facility*: SOAR (Goodman, SOI), William Herschel Telescope (ACAM), New Technology Telescope (EFOSC2). *Software*: Aladin Sky Atlas, Astropy (Astropy Collaboration et al. 2013), IDL, IRAF, SAOImage DS9.

**ORCID iDs**

Nora L. Eisner https://orcid.org/0000-0002-9138-9028
Matthew M. Knight https://orcid.org/0000-0003-2781-6897
Michael S. Kelley https://orcid.org/0000-0002-6702-7676
Colin Snodgrass https://orcid.org/0000-0001-9328-2905
Alan Fitzsimmons https://orcid.org/0000-0003-0250-9911
Rosita Kokotanekova https://orcid.org/0000-0003-4617-8878

**References**

A'Hearn, M. F., Millis, R. L., Schleicher, D. G., Osip, D. J., & Birch, P. V. 1995, Icarus, 118, 223
Bailey, M.E., Chambers, J.E., Hahn, G. 1992, A&A 257, 315-322
Bannister, M. T., Schwamb, M. E., Fraser, W. C., et al. 2017, ApJL, 851(2), L38
Battams, K. & Knight, M.M. 2017, Phil. Trans. A, 375, 20160257
Battams, K. & Lui, L. 2013, Central Bureau Electronic Telegrams 3631
Benn, C., Dee, K., & Agócs, T. 2008, Proc., SPIE, 7014
Bolin, B. T., Weaver, H. A., Fernandez, Y. R., et al. 2018, ApJL, 852, L2
Bonnarel, F., Ziaeepour, H., Bartlett, J. G., et al. 2000, A&AS, 143, 33
Bramich, D. M. 2008, MNRAS, 386, L77
Bramich, D. M., Horne, K., Albrow, M. D., et al. 2013, MNRAS, 428, 2275
Buzzoni, B., Delabre, B., Dekker, H., et al. 1984, Msngr, 38, 9
Carusi, A., Kresak, L., Perozzi, E., Vasecchi, G.B. 1987, A&A 187, 899-905.
Clemens, J. C., Crain, J. A., & Anderson, R. 2004, Proc. SPIE, 5492, 331
Chambers, K. C. 2017, VizieR Online Data Catalog, 2349

Colina, L., Bohlin, R. C., & Castelli, F. 1996, AJ, 112, 307

Combi, M. R., Boyd, Z., Lee, Y., et al. 2011, Icarus, 216(2), 449-461

Ćuk, M. 2018, ApJL, *852*(1), L15

Combi, M.R., Makinen, T.T., Bertaux, J.-L., Quemerais, E. & Ferron, S. 2019, Icarus 317, 610-620

Drahus, M., Guzik, P., Waniak, W., et al. 2018, Nature Astronomy, *2*(5), 407

Drahus, M., Kueppers, M., & Jarchow, C. et al. 2010, A&A, 510, A55

Dykhuis, M.J., Samarasinha, N.H., Mueller, B.E.A., & Storm, S.P. 2012, AAS DPS meeting #44, 314.10

Eisner, N., Knight, M.M., Battams, K., Kelley, M.S., Snodgrass, C. 2018, COSPAR B0.3-0017-18

Eisner, N., Knight, M. M., & Schleicher, D. G. 2017, AJ, 154, 196

Fitzsimmons, A., Snodgrass, C., Rozitis, B., et al. 2018, Nature Astronomy, *2*(2), 133

Fraser, W. C., Pravec, P., Fitzsimmons, A., et al. 2018, Nature Astronomy, *2*(5), 383

Fukugita, M., Ichikawa, T., Gunn, J. E., et al. 1996, AJ, 111, 1748

Granvik, M., Morbidelli, A., Jedicke, R., et al. 2016, Nature 530, 303-306

Green, D.W.E., Rickman, H., Porter, A.C. & Meech, K.J. 1990, Science 247, 1063-1067

Grynko, Y., Jockers, K., & Schwenn, R. 2004, A&A, 427(2), 755-761

Hainaut, O. R., Boehnhardt, H., & Protopapa, S. 2012, A & A, 546, A115

Harris, A.W., Young, J.W., Scaltriti, F., & Zappala, V. 1984, Icarus, 57, 251-258

Jewitt, D. 2018, arXiv preprint arXiv:1808.04885

Jewitt, D. 1991, in Iau Colloq. 116, Comets in the Post-Halley Era, ed. R. L. Newburn, Jr., M

Jewitt, D., Luu, J., Rajagopal, J., et al. 2017, ApJL, 850(2), L36

Jones, G.H., Knight, M.M., Battams, K., et al. 2018, Space Sci. Rev., 214, 20 (86pp.)

Knight, M.M. 2008, Ph.D. thesis, University of Maryland, College Park

Knight, M. M., Farnham, T. L., Schleicher, D. G., & Schwieterman, E. W. 2011, AJ, 141, 2

Knight, M.M., Fitzsimmons, A., Kelley, M.S.P. & Snodgrass, C. 2016, ApJL 823, L6 (6 pp)

Knight, M. M., Protopapa, S., Kelley, et al. 2017, ApJL, 851(2), L31

Knight, M. M., Schleicher D. G., Farnham T. L., Schwieterman E. W. & Christensen S. R. 2012 AJ 144 153

Knight, M.M., Walsh, K.J. 2013, ApJ, 776, L5

Kokotanekova, R., Snodgrass, C., Lacerda, P., et al. 2017, MNRAS, 471(3), 2974-3007

Kokotanekova, R., Snodgrass, C., Lacerda, P., et al. 2018, *MNRAS*, 479 (4), 4665-4680

Lamy, P., Faury, G., Llebaria, A., et al. 2013, Icarus 226, 1350-1398

Lamy, P., & Toth, I. 2009, Icarus, 201(2), 674-713

Lamy, P. L., Toth, I., Fernandez, Y. R., & Weaver, H. A. 2004, in Comets II, ed. M. C. Festou, H. U. Keller, & H. A. Weaver (Tucson: Univ. Arizona Press), 223

Lamy, P. L., Toth, I., Weaver, H. A., A'Hearn, M. F., & Jorda, L. 2009, A&A, 508, 1045

Langland-Shula, L.E. & Smith, G.H. 2007 ApJ 664, L119-L122

Levison, H. F., & Dones, L. 2014. Comet populations and cometary dynamics. In Encyclopedia of the Solar System, Third Edition, p. 708

Licandro, J., Tancredi, G., Lindgren, M., Rickman, H. & Hutton, R.G. 2000, Icarus 147, 161-179.

Lupton, R., 2005, http://www.sdss.org/dr5/algorithms/sdssUBVRITransform.html

Lowry, S., Duddy, S. R., Rozitis, B., et al. 2012, A&A, 548, A12

Mann, I., Kimura, H., Biesecker, D.A., et al., 2004, Space Sci. Rev. 110, 269-305

Marsden, B.G. 2005, Ann. Rev. Astron. Astrophys. 43, 75-102

Meech, K. 1996, Asteroids, Comets, and Meteors 1996 meeting. Available on line at http://www.ifa.hawaii.edu/~meech/papers/acm96.pdf

Meech, K. J., Hainaut, O. R., & Marsden, B. G. 2004, Icarus, 170(2), 463-491

Meech K. J., Weryk R., Micheli M. et al., 2017 Nature 552 378

McIntosh, B.A. 1990, Icarus 86, 299-304

Micheli, M., Farnocchia, D., Meech, K. J., et al., 2018, Nature, 559, 223

Mueller, B.E.A. & Samarasinha, N. 2018, AJ 156, 107 (7pp)

Nolan, M.C., Harmon, J.K., Howell, E.S., et al. 2006, BAAS 38, 504

Ohtsuka, K., Nakano, S. & Yoshikawa, M. 2003, PASJ 55, 321-324

Reddy, V., Dunn, T. L., Thomas, C. A., Moskovitz, N. A., & Burbine, T. H., 2015, Asteroids IV, 43

Russell, H. N. 1916, ApJ, 43, 173

Samarasinha, N. H., Mueller, B. E. A., Belton, M. J. S., & Jorda, L. 2004, in Comets II, ed. M. C. Festou, H. U. Keller, & H. A. Weaver (Tucson, AZ: Univ. Arizona Press), 281

Schleicher, D.G. 2008, AJ 136, 2204-2213

Sekanina, Z. 1988, AJ 96, 1455-1475

Sekanina, Z. 1990, AJ 99, 1268-1277

Sekanina, Z. 1991, J. Roy. Astron. Soc. Can., 85, 324-376

Sekanina, Z. 2013, Central Bureau Electronic Telegrams 3642

Sekanina, Z. & Chodas, P.W. 2005, ApJSS 161, 551-586

Smith, J. A., Tucker, D. L., Kent, S., et al. 2002, AJ, 123, 2121

Snodgrass, C., Fitzsimmons, A., & Lowry, S. C. 2005, A&A, 444, 287

Snodgrass C., Jehin E., Manfoid J. et al. 2016, A&A, 588, A80

Snodgrass C., Tubiana C., Bramich D. M. et al. 2013, A&A, 557, A33

Stellingwerf R. F. 1978, ApJ, 224, 953

Tonry, J. L., Stubbs, C. W., Lykkem, et al., 2012, ApJ, 750, 99

Ye, Q. Z., Zhang, Q., Kelley, M. S., & Brown, P. G. 2017, ApJL, 851(1), L5

Walker, A. R., Boccas, M., Bonati, M., al. 2003, Proc. SPIE, 4841, 286

## Supplementary Table –
## All observed $r'$-band Magnitudes and their Statistical Uncertainties.

| Date[a] | UT[b] | $m_{r'}$[c] | $\sigma_{r'}$[d] | Date[a] | UT[b] | $m_{r'}$[c] | $\sigma_{r'}$[d] | Date[a] | UT[b] | $m_{r'}$[c] | $\sigma_{r'}$[d] |
|---|---|---|---|---|---|---|---|---|---|---|---|
| 2017 Apr 7 | 08:22:04 | 20.458 | 0.053 | 2017 Jul 2 | 04:05:57 | 18.498 | 0.005 | 2017 Jul 2 | 05:35:27 | 18.602 | 0.004 |
| 2017 Apr 7 | 08:23:09 | 20.489 | 0.022 | 2017 Jul 2 | 04:08:06 | 18.450 | 0.004 | 2017 Jul 2 | 05:37:36 | 18.584 | 0.004 |
| 2017 Apr 7 | 08:27:05 | 20.474 | 0.022 | 2017 Jul 2 | 04:10:15 | 18.423 | 0.004 | 2017 Jul 2 | 05:39:45 | 18.526 | 0.004 |
| 2017 Apr 7 | 08:34:17 | 20.467 | 0.021 | 2017 Jul 2 | 04:12:23 | 18.382 | 0.004 | 2017 Jul 2 | 05:41:53 | 18.536 | 0.004 |
| 2017 Apr 7 | 09:00:45 | 20.465 | 0.021 | 2017 Jul 2 | 04:14:32 | 18.346 | 0.004 | 2017 Jul 2 | 05:44:02 | 18.524 | 0.004 |
| 2017 Apr 7 | 09:11:27 | 20.388 | 0.019 | 2017 Jul 2 | 04:16:41 | 18.320 | 0.004 | 2017 Jul 2 | 05:46:11 | 18.487 | 0.004 |
| 2017 Apr 7 | 09:37:31 | 20.190 | 0.015 | 2017 Jul 2 | 04:18:50 | 18.328 | 0.004 | 2017 Jul 2 | 05:48:20 | 18.467 | 0.003 |
| 2017 Jul 2 | 02:48:44 | 18.465 | 0.005 | 2017 Jul 2 | 04:21:32 | 18.267 | 0.004 | 2017 Jul 2 | 05:51:17 | 18.428 | 0.003 |
| 2017 Jul 2 | 02:50:53 | 18.450 | 0.005 | 2017 Jul 2 | 04:23:41 | 18.255 | 0.004 | 2017 Jul 2 | 05:53:26 | 18.399 | 0.003 |
| 2017 Jul 2 | 02:54:26 | 18.405 | 0.005 | 2017 Jul 2 | 04:25:50 | 18.221 | 0.004 | 2017 Jul 2 | 05:55:35 | 18.386 | 0.003 |
| 2017 Jul 2 | 02:56:35 | 18.393 | 0.005 | 2017 Jul 2 | 04:27:59 | 18.208 | 0.004 | 2017 Jul 2 | 05:57:44 | 18.380 | 0.003 |
| 2017 Jul 2 | 02:58:44 | 18.395 | 0.005 | 2017 Jul 2 | 04:30:08 | 18.207 | 0.004 | 2017 Jul 2 | 05:59:53 | 18.368 | 0.003 |
| 2017 Jul 2 | 03:00:53 | 18.439 | 0.005 | 2017 Jul 2 | 04:32:17 | 18.203 | 0.004 | 2017 Jul 2 | 06:02:02 | 18.358 | 0.003 |
| 2017 Jul 2 | 03:03:02 | 18.423 | 0.005 | 2017 Jul 2 | 04:34:26 | 18.192 | 0.003 | 2017 Jul 2 | 06:04:11 | 18.388 | 0.003 |
| 2017 Jul 2 | 03:05:11 | 18.391 | 0.005 | 2017 Jul 2 | 04:36:35 | 18.210 | 0.004 | 2017 Jul 2 | 06:06:20 | 18.369 | 0.003 |
| 2017 Jul 2 | 03:07:20 | 18.349 | 0.005 | 2017 Jul 2 | 04:38:44 | 18.208 | 0.003 | 2017 Jul 2 | 06:08:29 | 18.372 | 0.003 |
| 2017 Jul 2 | 03:09:30 | 18.372 | 0.005 | 2017 Jul 2 | 04:42:14 | 18.210 | 0.003 | 2017 Jul 2 | 06:12:02 | 18.360 | 0.003 |
| 2017 Jul 2 | 03:11:39 | 18.366 | 0.005 | 2017 Jul 2 | 04:44:23 | 18.219 | 0.003 | 2017 Jul 2 | 06:14:12 | 18.367 | 0.003 |
| 2017 Jul 2 | 03:14:12 | 18.386 | 0.005 | 2017 Jul 2 | 04:46:33 | 18.230 | 0.003 | 2017 Jul 2 | 06:16:21 | 18.369 | 0.003 |
| 2017 Jul 2 | 03:20:39 | 18.422 | 0.005 | 2017 Jul 2 | 04:48:42 | 18.248 | 0.004 | 2017 Jul 2 | 06:18:30 | 18.376 | 0.003 |
| 2017 Jul 2 | 03:22:47 | 18.451 | 0.005 | 2017 Jul 2 | 04:50:51 | 18.273 | 0.004 | 2017 Jul 2 | 06:20:39 | 18.373 | 0.003 |
| 2017 Jul 2 | 03:24:56 | 18.490 | 0.005 | 2017 Jul 2 | 04:53:00 | 18.289 | 0.004 | 2017 Jul 2 | 06:22:48 | 18.367 | 0.003 |
| 2017 Jul 2 | 03:27:05 | 18.484 | 0.005 | 2017 Jul 2 | 04:55:09 | 18.327 | 0.004 | 2017 Jul 2 | 06:24:56 | 18.365 | 0.003 |
| 2017 Jul 2 | 03:29:14 | 18.499 | 0.005 | 2017 Jul 2 | 04:57:17 | 18.359 | 0.004 | 2017 Jul 2 | 06:27:05 | 18.382 | 0.003 |
| 2017 Jul 2 | 03:31:23 | 18.499 | 0.005 | 2017 Jul 2 | 04:59:26 | 18.377 | 0.004 | 2017 Jul 2 | 06:29:14 | 18.402 | 0.003 |
| 2017 Jul 2 | 03:33:56 | 18.536 | 0.005 | 2017 Jul 2 | 05:11:26 | 18.515 | 0.004 | 2017 Jul 2 | 06:44:38 | 18.400 | 0.003 |
| 2017 Jul 2 | 03:36:05 | 18.537 | 0.005 | 2017 Jul 2 | 05:13:35 | 18.531 | 0.004 | 2017 Jul 2 | 06:46:47 | 18.413 | 0.003 |
| 2017 Jul 2 | 03:38:14 | 18.566 | 0.006 | 2017 Jul 2 | 05:15:44 | 18.469 | 0.004 | 2017 Jul 2 | 06:48:56 | 18.419 | 0.003 |
| 2017 Jul 2 | 03:40:23 | 18.578 | 0.006 | 2017 Jul 2 | 05:17:53 | 18.598 | 0.004 | 2017 Jul 2 | 06:51:05 | 18.425 | 0.003 |
| 2017 Jul 2 | 03:42:32 | 18.581 | 0.005 | 2017 Jul 2 | 05:20:02 | 18.638 | 0.005 | 2017 Jul 2 | 06:53:14 | 18.405 | 0.003 |
| 2017 Jul 2 | 03:44:41 | 18.602 | 0.005 | 2017 Jul 2 | 05:22:11 | 18.645 | 0.005 | 2017 Jul 2 | 06:55:23 | 18.421 | 0.003 |
| 2017 Jul 2 | 03:46:50 | 18.589 | 0.005 | 2017 Jul 2 | 05:24:20 | 18.657 | 0.004 | 2017 Jul 2 | 06:57:33 | 18.393 | 0.003 |
| 2017 Jul 2 | 03:48:59 | 18.578 | 0.005 | 2017 Jul 2 | 05:26:29 | 18.652 | 0.004 | 2017 Jul 2 | 06:59:42 | 18.388 | 0.003 |
| 2017 Jul 2 | 03:51:08 | 18.587 | 0.005 | 2017 Jul 2 | 05:28:38 | 18.658 | 0.004 | 2017 Jul 2 | 07:01:51 | 18.380 | 0.003 |
| 2017 Jul 2 | 04:01:38 | 18.499 | 0.005 | 2017 Jul 2 | 05:31:09 | 18.646 | 0.004 | 2017 Jul 2 | 07:04:36 | 18.351 | 0.003 |
| 2017 Jul 2 | 04:03:48 | 18.486 | 0.005 | 2017 Jul 2 | 05:33:18 | 18.629 | 0.004 | 2017 Jul 2 | 07:06:44 | 18.354 | 0.003 |

Supplementary Table - Continued

| Date[a] | UT[b] | $m_r$[c] | $\sigma_r$[d] | Date[a] | UT[b] | $m_r$[c] | $\sigma_r$[d] | Date[a] | UT[b] | $m_r$[c] | $\sigma_r$[d] |
|---|---|---|---|---|---|---|---|---|---|---|---|
| 2017 Jul 2 | 07:08:53 | 18.353 | 0.003 | 2017 Jul 2 | 08:52:11 | 18.234 | 0.003 | 2017 Jul 3 | 09:08:53 | 18.141 | 0.004 |
| 2017 Jul 2 | 07:11:02 | 18.380 | 0.003 | 2017 Jul 2 | 08:54:20 | 18.255 | 0.003 | 2017 Jul 3 | 09:11:02 | 18.114 | 0.004 |
| 2017 Jul 2 | 07:13:11 | 18.373 | 0.003 | 2017 Jul 2 | 08:56:29 | 18.269 | 0.003 | 2018 Jun 25 | 00:19:09 | 19.203 | 0.035 |
| 2017 Jul 2 | 07:15:20 | 18.351 | 0.003 | 2017 Jul 2 | 08:58:38 | 18.289 | 0.003 | 2018 Jun 25 | 00:51:53 | 19.151 | 0.019 |
| 2017 Jul 2 | 07:17:29 | 18.339 | 0.003 | 2017 Jul 2 | 09:00:47 | 18.327 | 0.003 | 2018 Jun 25 | 01:20:43 | 19.259 | 0.029 |
| 2017 Jul 2 | 07:19:38 | 18.363 | 0.003 | 2017 Jul 2 | 09:10:57 | 18.494 | 0.003 | 2018 Jun 25 | 01:49:48 | 19.303 | 0.022 |
| 2017 Jul 2 | 07:21:47 | 18.391 | 0.003 | 2017 Jul 2 | 09:13:06 | 18.507 | 0.004 | 2018 Jun 25 | 01:52:54 | 19.320 | 0.019 |
| 2017 Jul 2 | 07:24:26 | 18.393 | 0.003 | 2017 Jul 2 | 09:15:15 | 18.530 | 0.004 | 2018 Jun 25 | 02:03:01 | 19.472 | 0.025 |
| 2017 Jul 2 | 07:26:35 | 18.403 | 0.003 | 2017 Jul 2 | 09:17:24 | 18.476 | 0.003 | 2018 Jun 25 | 02:09:12 | 19.248 | 0.021 |
| 2017 Jul 2 | 07:28:44 | 18.415 | 0.003 | 2017 Jul 2 | 09:21:41 | 18.518 | 0.003 | 2018 Jun 25 | 02:12:19 | 19.205 | 0.020 |
| 2017 Jul 2 | 07:30:53 | 18.418 | 0.003 | 2017 Jul 2 | 09:23:50 | 18.642 | 0.004 | 2018 Jun 25 | 02:22:01 | 19.236 | 0.018 |
| 2017 Jul 2 | 07:33:02 | 18.421 | 0.003 | 2017 Jul 2 | 09:25:59 | 18.618 | 0.004 | 2018 Jun 25 | 02:25:07 | 19.194 | 0.016 |
| 2017 Jul 2 | 07:35:11 | 18.436 | 0.003 | 2017 Jul 2 | 09:28:08 | 18.662 | 0.004 | 2018 Jun 25 | 02:28:12 | 19.062 | 0.015 |
| 2017 Jul 2 | 07:39:29 | 18.495 | 0.003 | 2017 Jul 2 | 09:31:29 | 18.680 | 0.004 | 2018 Jun 25 | 02:31:18 | 19.025 | 0.013 |
| 2017 Jul 2 | 07:41:39 | 18.520 | 0.003 | 2017 Jul 2 | 09:33:38 | 18.698 | 0.004 | 2018 Jun 25 | 02:35:15 | 19.221 | 0.014 |
| 2017 Jul 2 | 07:52:27 | 18.589 | 0.003 | 2017 Jul 2 | 09:35:47 | 18.663 | 0.004 | 2018 Jun 25 | 02:38:20 | 19.220 | 0.014 |
| 2017 Jul 2 | 07:54:35 | 18.615 | 0.004 | 2017 Jul 2 | 10:28:54 | 18.344 | 0.003 | 2018 Jun 25 | 02:41:26 | 19.180 | 0.014 |
| 2017 Jul 2 | 07:56:44 | 18.609 | 0.004 | 2017 Jul 2 | 10:31:27 | 18.317 | 0.003 | 2018 Jun 25 | 02:44:32 | 19.200 | 0.014 |
| 2017 Jul 2 | 07:58:53 | 18.605 | 0.004 | 2017 Jul 2 | 10:34:08 | 18.364 | 0.004 | 2018 Jun 25 | 02:47:38 | 19.153 | 0.014 |
| 2017 Jul 2 | 08:01:02 | 18.596 | 0.004 | 2017 Jul 2 | 10:36:18 | 18.375 | 0.005 | 2018 Jun 25 | 02:52:13 | 19.179 | 0.012 |
| 2017 Jul 2 | 08:03:11 | 18.614 | 0.004 | 2017 Jul 2 | 10:38:27 | 18.368 | 0.006 | 2018 Jun 25 | 02:55:19 | 19.199 | 0.013 |
| 2017 Jul 2 | 08:05:20 | 18.600 | 0.003 | 2017 Jul 2 | 10:41:06 | 18.330 | 0.006 | 2018 Jun 25 | 02:58:25 | 19.225 | 0.014 |
| 2017 Jul 2 | 08:07:29 | 18.578 | 0.003 | 2017 Jul 2 | 10:43:14 | 18.316 | 0.007 | 2018 Jun 25 | 03:01:31 | 19.197 | 0.013 |
| 2017 Jul 2 | 08:09:38 | 18.580 | 0.003 | 2017 Jul 2 | 10:45:23 | 18.342 | 0.008 | 2018 Jun 25 | 03:04:37 | 19.108 | 0.012 |
| 2017 Jul 2 | 08:21:04 | 18.371 | 0.003 | 2017 Jul 3 | 08:38:15 | 18.604 | 0.012 | 2018 Jun 25 | 03:25:25 | 19.248 | 0.016 |
| 2017 Jul 2 | 08:23:42 | 18.329 | 0.003 | 2017 Jul 3 | 08:41:36 | 18.526 | 0.003 | 2018 Jun 25 | 03:31:36 | 19.293 | 0.016 |
| 2017 Jul 2 | 08:25:51 | 18.307 | 0.003 | 2017 Jul 3 | 08:43:45 | 18.499 | 0.003 | 2018 Jun 25 | 03:34:42 | 19.343 | 0.018 |
| 2017 Jul 2 | 08:28:00 | 18.276 | 0.003 | 2017 Jul 3 | 08:45:54 | 18.493 | 0.003 | 2018 Jun 25 | 03:38:19 | 19.364 | 0.018 |
| 2017 Jul 2 | 08:30:09 | 18.285 | 0.003 | 2017 Jul 3 | 08:48:03 | 18.457 | 0.003 | 2018 Jun 25 | 04:00:19 | 19.229 | 0.017 |
| 2017 Jul 2 | 08:32:18 | 18.272 | 0.003 | 2017 Jul 3 | 08:50:56 | 18.385 | 0.003 | 2018 Jun 25 | 04:06:31 | 19.140 | 0.015 |
| 2017 Jul 2 | 08:34:27 | 18.255 | 0.003 | 2017 Jul 3 | 08:53:05 | 18.363 | 0.003 | 2018 Jun 25 | 04:09:51 | 19.063 | 0.016 |
| 2017 Jul 2 | 08:36:35 | 18.243 | 0.003 | 2017 Jul 3 | 08:55:14 | 18.358 | 0.003 | 2018 Jun 25 | 04:16:02 | 19.315 | 0.018 |
| 2017 Jul 2 | 08:38:44 | 18.223 | 0.003 | 2017 Jul 3 | 08:57:23 | 18.316 | 0.003 | 2018 Jun 25 | 04:19:09 | 19.310 | 0.021 |
| 2017 Jul 2 | 08:40:53 | 18.230 | 0.003 | 2017 Jul 3 | 08:59:32 | 18.279 | 0.003 | 2018 Jun 25 | 05:37:43 | 19.410 | 0.030 |
| 2017 Jul 2 | 08:43:35 | 18.200 | 0.003 | 2017 Jul 3 | 09:02:26 | 18.211 | 0.003 | 2018 Jun 25 | 05:40:48 | 19.280 | 0.032 |
| 2017 Jul 2 | 08:45:44 | 18.184 | 0.003 | 2017 Jul 3 | 09:04:35 | 18.221 | 0.003 | 2018 Jun 25 | 05:43:55 | 19.139 | 0.100 |
| 2017 Jul 2 | 08:50:02 | 18.211 | 0.003 | 2017 Jul 3 | 09:06:44 | 18.167 | 0.003 | 2018 Jun 25 | 05:47:01 | 19.155 | 0.037 |

| Date[a] | UT[b] | $m_r$[c] | $\sigma_r$[d] | Date[a] | UT[b] | $m_r$[c] | $\sigma_r$[d] | Date[a] | UT[b] | $m_r$[c] | $\sigma_r$[d] |
|---|---|---|---|---|---|---|---|---|---|---|---|
| 2018 Jun 25 | 06:06:10 | 19.276 | 0.036 | 2018 Jul 10 | 23:29:06 | 19.795 | 0.012 | 2018 Jul 11 | 00:10:57 | 19.863 | 0.024 |
| 2018 Jun 25 | 06:15:56 | 19.212 | 0.014 | 2018 Jul 10 | 23:30:12 | 19.802 | 0.014 | 2018 Jul 11 | 00:12:02 | 19.896 | 0.022 |
| 2018 Jun 25 | 06:19:01 | 19.294 | 0.017 | 2018 Jul 10 | 23:31:18 | 19.801 | 0.014 | 2018 Jul 11 | 00:13:08 | 19.877 | 0.019 |
| 2018 Jun 25 | 06:22:07 | 19.264 | 0.015 | 2018 Jul 10 | 23:32:24 | 19.804 | 0.014 | 2018 Jul 11 | 00:14:14 | 19.872 | 0.018 |
| 2018 Jun 25 | 06:40:06 | 19.241 | 0.015 | 2018 Jul 10 | 23:33:30 | 19.807 | 0.014 | 2018 Jul 11 | 00:18:38 | 19.965 | 0.017 |
| 2018 Jun 25 | 06:46:18 | 19.275 | 0.015 | 2018 Jul 10 | 23:35:12 | 19.812 | 0.014 | 2018 Jul 11 | 00:19:48 | 19.958 | 0.017 |
| 2018 Jun 25 | 06:52:44 | 19.233 | 0.014 | 2018 Jul 10 | 23:36:17 | 19.813 | 0.014 | 2018 Jul 11 | 00:28:34 | 19.964 | 0.028 |
| 2018 Jun 25 | 06:55:50 | 19.227 | 0.013 | 2018 Jul 10 | 23:37:23 | 19.803 | 0.014 | 2018 Jul 11 | 00:29:40 | 19.978 | 0.027 |
| 2018 Jun 25 | 06:58:56 | 19.221 | 0.015 | 2018 Jul 10 | 23:38:29 | 19.809 | 0.015 | 2018 Jul 11 | 00:30:59 | 19.956 | 0.022 |
| 2018 Jun 25 | 07:02:02 | 19.180 | 0.013 | 2018 Jul 10 | 23:39:35 | 19.817 | 0.014 | 2018 Jul 11 | 00:32:05 | 19.964 | 0.017 |
| 2018 Jun 25 | 07:05:08 | 19.128 | 0.012 | 2018 Jul 10 | 23:40:41 | 19.821 | 0.015 | 2018 Jul 11 | 00:33:10 | 19.921 | 0.014 |
| 2018 Jun 25 | 07:08:39 | 19.254 | 0.014 | 2018 Jul 10 | 23:41:47 | 19.812 | 0.014 | 2018 Jul 11 | 00:34:17 | 19.983 | 0.020 |
| 2018 Jun 25 | 07:21:03 | 19.191 | 0.015 | 2018 Jul 10 | 23:42:52 | 19.797 | 0.020 | 2018 Jul 11 | 00:35:22 | 19.970 | 0.016 |
| 2018 Jun 25 | 07:33:04 | 19.471 | 0.019 | 2018 Jul 10 | 23:43:59 | 19.816 | 0.022 | 2018 Jul 11 | 00:36:28 | 19.934 | 0.027 |
| 2018 Jun 25 | 07:36:10 | 19.246 | 0.018 | 2018 Jul 10 | 23:45:04 | 19.818 | 0.014 | 2018 Jul 11 | 00:37:35 | 19.945 | 0.019 |
| 2018 Jun 25 | 07:39:16 | 19.272 | 0.017 | 2018 Jul 10 | 23:46:27 | 19.807 | 0.016 | 2018 Jul 11 | 00:38:41 | 19.953 | 0.016 |
| 2018 Jun 25 | 07:42:22 | 19.424 | 0.019 | 2018 Jul 10 | 23:47:33 | 19.830 | 0.014 | 2018 Jul 11 | 00:39:47 | 19.935 | 0.014 |
| 2018 Jun 25 | 07:58:26 | 19.313 | 0.022 | 2018 Jul 10 | 23:48:39 | 19.807 | 0.020 | 2018 Jul 11 | 00:40:53 | 19.986 | 0.013 |
| 2018 Jun 25 | 08:11:25 | 19.363 | 0.019 | 2018 Jul 10 | 23:49:45 | 19.814 | 0.016 | 2018 Jul 11 | 00:42:08 | 19.976 | 0.017 |
| 2018 Jun 25 | 08:35:14 | 19.199 | 0.014 | 2018 Jul 10 | 23:50:51 | 19.801 | 0.039 | 2018 Jul 11 | 00:44:19 | 19.954 | 0.034 |
| 2018 Jun 25 | 08:38:19 | 19.194 | 0.015 | 2018 Jul 10 | 23:51:57 | 19.793 | 0.015 | 2018 Jul 11 | 00:45:25 | 19.946 | 0.014 |
| 2018 Jun 25 | 08:41:25 | 19.270 | 0.015 | 2018 Jul 10 | 23:53:03 | 19.795 | 0.023 | 2018 Jul 11 | 00:46:31 | 19.953 | 0.013 |
| 2018 Jun 25 | 08:44:31 | 19.300 | 0.015 | 2018 Jul 10 | 23:54:09 | 19.808 | 0.022 | 2018 Jul 11 | 00:47:37 | 19.964 | 0.013 |
| 2018 Jun 25 | 08:47:37 | 19.215 | 0.012 | 2018 Jul 10 | 23:55:15 | 19.814 | 0.028 | 2018 Jul 11 | 00:48:43 | 19.994 | 0.013 |
| 2018 Jun 25 | 08:50:56 | 19.216 | 0.014 | 2018 Jul 10 | 23:56:21 | 19.817 | 0.038 | 2018 Jul 11 | 00:49:49 | 20.001 | 0.015 |
| 2018 Jun 25 | 08:54:02 | 19.259 | 0.014 | 2018 Jul 10 | 23:57:36 | 19.807 | 0.026 | 2018 Jul 11 | 00:50:55 | 19.959 | 0.012 |
| 2018 Jun 25 | 08:57:08 | 19.242 | 0.016 | 2018 Jul 10 | 23:58:42 | 19.786 | 0.017 | 2018 Jul 11 | 00:54:17 | 19.907 | 0.012 |
| 2018 Jun 25 | 09:06:46 | 19.197 | 0.015 | 2018 Jul 10 | 23:59:48 | 19.818 | 0.020 | 2018 Jul 11 | 00:55:23 | 19.936 | 0.012 |
| 2018 Jun 25 | 09:12:57 | 19.243 | 0.014 | 2018 Jul 11 | 00:00:54 | 19.802 | 0.016 | 2018 Jul 11 | 00:56:29 | 19.920 | 0.013 |
| 2018 Jun 25 | 09:16:04 | 19.175 | 0.013 | 2018 Jul 11 | 00:02:00 | 19.805 | 0.020 | 2018 Jul 11 | 00:57:35 | 19.943 | 0.013 |
| 2018 Jun 25 | 09:19:10 | 19.279 | 0.015 | 2018 Jul 11 | 00:03:05 | 19.816 | 0.020 | 2018 Jul 11 | 00:58:40 | 19.914 | 0.012 |
| 2018 Jul 10 | 23:21:28 | 19.797 | 0.014 | 2018 Jul 11 | 00:04:11 | 19.841 | 0.024 | 2018 Jul 11 | 00:59:46 | 19.888 | 0.013 |
| 2018 Jul 10 | 23:22:46 | 19.786 | 0.012 | 2018 Jul 11 | 00:05:17 | 19.844 | 0.018 | 2018 Jul 11 | 01:00:52 | 19.863 | 0.011 |
| 2018 Jul 10 | 23:23:52 | 19.781 | 0.014 | 2018 Jul 11 | 00:06:23 | 19.817 | 0.035 | 2018 Jul 11 | 01:01:58 | 19.910 | 0.012 |
| 2018 Jul 10 | 23:24:58 | 19.794 | 0.013 | 2018 Jul 11 | 00:07:29 | 19.827 | 0.016 | 2018 Jul 11 | 01:03:04 | 19.918 | 0.013 |
| 2018 Jul 10 | 23:26:04 | 19.796 | 0.013 | 2018 Jul 11 | 00:08:45 | 19.858 | 0.018 | 2018 Jul 11 | 01:06:29 | 19.880 | 0.019 |
| 2018 Jul 10 | 23:27:09 | 19.792 | 0.012 | 2018 Jul 11 | 00:09:51 | 19.850 | 0.019 | 2018 Jul 11 | 01:09:13 | 19.857 | 0.012 |

| Date[a] | UT[b] | $m_r$[c] | $\sigma_r$[d] | Date[a] | UT[b] | $m_r$[c] | $\sigma_r$[d] | Date[a] | UT[b] | $m_r$[c] | $\sigma_r$[d] |
|---|---|---|---|---|---|---|---|---|---|---|---|
| 2018 Jul 11 | 01:12:14 | 19.850 | 0.013 | 2018 Jul 11 | 01:58:43 | 19.873 | 0.020 | 2018 Jul 11 | 02:23:53 | 20.037 | 0.023 |
| 2018 Jul 11 | 01:15:13 | 19.831 | 0.020 | 2018 Jul 11 | 01:59:19 | 19.828 | 0.018 | 2018 Jul 11 | 02:24:29 | 20.010 | 0.023 |
| 2018 Jul 11 | 01:16:14 | 19.803 | 0.022 | 2018 Jul 11 | 01:59:55 | 19.881 | 0.020 | 2018 Jul 11 | 02:25:05 | 20.001 | 0.023 |
| 2018 Jul 11 | 01:16:49 | 19.812 | 0.028 | 2018 Jul 11 | 02:00:31 | 19.881 | 0.021 | 2018 Jul 11 | 02:25:41 | 19.998 | 0.023 |
| 2018 Jul 11 | 01:17:26 | 19.808 | 0.036 | 2018 Jul 11 | 02:01:08 | 19.867 | 0.020 | 2018 Jul 11 | 02:26:17 | 19.991 | 0.023 |
| 2018 Jul 11 | 01:18:01 | 19.825 | 0.032 | 2018 Jul 11 | 02:01:44 | 19.889 | 0.020 | 2018 Jul 11 | 02:26:52 | 20.017 | 0.023 |
| 2018 Jul 11 | 01:18:37 | 19.814 | 0.030 | 2018 Jul 11 | 02:02:19 | 19.890 | 0.019 | 2018 Jul 11 | 02:27:28 | 19.996 | 0.024 |
| 2018 Jul 11 | 01:19:14 | 19.812 | 0.022 | 2018 Jul 11 | 02:02:55 | 19.918 | 0.019 | 2018 Jul 11 | 02:47:04 | 19.991 | 0.028 |
| 2018 Jul 11 | 01:19:50 | 19.793 | 0.021 | 2018 Jul 11 | 02:03:31 | 19.929 | 0.018 | 2018 Jul 11 | 02:47:39 | 20.037 | 0.024 |
| 2018 Jul 11 | 01:20:26 | 19.789 | 0.016 | 2018 Jul 11 | 02:04:07 | 19.922 | 0.019 | 2018 Jul 11 | 02:48:15 | 19.983 | 0.025 |
| 2018 Jul 11 | 01:29:41 | 19.739 | 0.016 | 2018 Jul 11 | 02:07:28 | 19.900 | 0.019 | 2018 Jul 11 | 02:49:27 | 20.019 | 0.027 |
| 2018 Jul 11 | 01:30:17 | 19.764 | 0.016 | 2018 Jul 11 | 02:08:04 | 19.890 | 0.017 | 2018 Jul 11 | 02:50:03 | 20.006 | 0.027 |
| 2018 Jul 11 | 01:30:53 | 19.780 | 0.017 | 2018 Jul 11 | 02:08:40 | 19.881 | 0.019 | 2018 Jul 11 | 02:50:39 | 20.004 | 0.025 |
| 2018 Jul 11 | 01:31:29 | 19.793 | 0.016 | 2018 Jul 11 | 02:09:16 | 19.911 | 0.019 | 2018 Jul 11 | 02:51:15 | 19.982 | 0.029 |
| 2018 Jul 11 | 01:32:05 | 19.823 | 0.018 | 2018 Jul 11 | 02:09:52 | 19.947 | 0.019 | 2018 Jul 11 | 02:51:51 | 20.017 | 0.031 |
| 2018 Jul 11 | 01:32:41 | 19.804 | 0.018 | 2018 Jul 11 | 02:10:28 | 19.961 | 0.018 | 2018 Jul 11 | 02:52:27 | 20.001 | 0.032 |
| 2018 Jul 11 | 01:33:17 | 19.823 | 0.024 | 2018 Jul 11 | 02:11:04 | 19.917 | 0.020 | 2018 Jul 11 | 02:53:20 | 20.037 | 0.029 |
| 2018 Jul 11 | 01:33:52 | 19.772 | 0.018 | 2018 Jul 11 | 02:11:39 | 19.940 | 0.022 | 2018 Jul 11 | 02:53:56 | 19.952 | 0.035 |
| 2018 Jul 11 | 01:34:28 | 19.830 | 0.022 | 2018 Jul 11 | 02:12:15 | 19.980 | 0.020 | 2018 Jul 11 | 02:54:32 | 19.950 | 0.060 |
| 2018 Jul 11 | 01:35:04 | 19.812 | 0.017 | 2018 Jul 11 | 02:12:51 | 19.972 | 0.021 | 2018 Jul 11 | 02:55:07 | 19.955 | 0.068 |
| 2018 Jul 11 | 01:35:40 | 19.802 | 0.017 | 2018 Jul 11 | 02:13:27 | 19.944 | 0.021 | 2018 Jul 11 | 02:55:43 | 19.937 | 0.173 |
| 2018 Jul 11 | 01:36:16 | 19.768 | 0.017 | 2018 Jul 11 | 02:14:03 | 20.002 | 0.021 | 2018 Jul 11 | 02:56:19 | 19.948 | 0.182 |
| 2018 Jul 11 | 01:36:52 | 19.795 | 0.025 | 2018 Jul 11 | 02:14:39 | 19.965 | 0.022 | 2018 Jul 11 | 02:56:55 | 19.981 | 0.245 |
| 2018 Jul 11 | 01:37:28 | 19.792 | 0.032 | 2018 Jul 11 | 02:15:15 | 19.974 | 0.021 | 2018 Jul 11 | 02:57:31 | 19.980 | 0.203 |
| 2018 Jul 11 | 01:38:04 | 19.781 | 0.038 | 2018 Jul 11 | 02:15:50 | 20.016 | 0.023 | 2018 Jul 11 | 02:58:07 | 19.955 | 0.203 |
| 2018 Jul 11 | 01:38:39 | 19.784 | 0.052 | 2018 Jul 11 | 02:16:26 | 19.993 | 0.022 | 2018 Jul 11 | 02:58:43 | 19.940 | 0.193 |
| 2018 Jul 11 | 01:39:15 | 19.784 | 0.054 | 2018 Jul 11 | 02:17:02 | 20.042 | 0.022 | 2018 Jul 11 | 02:59:39 | 19.960 | 0.062 |
| 2018 Jul 11 | 01:39:51 | 19.797 | 0.043 | 2018 Jul 11 | 02:17:38 | 20.041 | 0.022 | 2018 Jul 11 | 03:00:45 | 19.949 | 0.030 |
| 2018 Jul 11 | 01:52:45 | 19.832 | 0.018 | 2018 Jul 11 | 02:18:14 | 19.995 | 0.021 | 2018 Jul 11 | 03:01:51 | 19.952 | 0.045 |
| 2018 Jul 11 | 01:53:21 | 19.837 | 0.018 | 2018 Jul 11 | 02:18:50 | 20.041 | 0.024 | 2018 Jul 11 | 03:02:56 | 19.925 | 0.036 |
| 2018 Jul 11 | 01:53:56 | 19.848 | 0.017 | 2018 Jul 11 | 02:19:42 | 19.986 | 0.020 | 2018 Jul 11 | 03:04:02 | 19.964 | 0.020 |
| 2018 Jul 11 | 01:55:08 | 19.850 | 0.019 | 2018 Jul 11 | 02:20:18 | 20.015 | 0.023 | 2018 Jul 11 | 03:09:32 | 19.918 | 0.018 |
| 2018 Jul 11 | 01:55:44 | 19.894 | 0.019 | 2018 Jul 11 | 02:20:54 | 20.038 | 0.023 | 2018 Jul 11 | 03:12:52 | 19.908 | 0.023 |
| 2018 Jul 11 | 01:56:20 | 19.881 | 0.019 | 2018 Jul 11 | 02:21:30 | 20.058 | 0.023 | 2018 Jul 11 | 03:15:45 | 19.914 | 0.069 |
| 2018 Jul 11 | 01:56:55 | 19.865 | 0.019 | 2018 Jul 11 | 02:22:06 | 20.044 | 0.023 | 2018 Jul 11 | 03:18:28 | 19.886 | 0.086 |
| 2018 Jul 11 | 01:57:31 | 19.844 | 0.018 | 2018 Jul 11 | 02:22:41 | 20.003 | 0.026 | 2018 Jul 11 | 03:19:44 | 19.898 | 0.081 |
| 2018 Jul 11 | 01:58:07 | 19.872 | 0.020 | 2018 Jul 11 | 02:23:17 | 20.009 | 0.023 | 2018 Jul 11 | 03:20:50 | 19.887 | 0.026 |

| Date[a] | UT[b] | $m_r$[c] | $\sigma_r$[d] | Date[a] | UT[b] | $m_r$[c] | $\sigma_r$[d] | Date[a] | UT[b] | $m_r$[c] | $\sigma_r$[d] |
|---|---|---|---|---|---|---|---|---|---|---|---|
| 2018 Jul 11 | 03:21:56 | 19.866 | 0.020 | 2018 Jul 11 | 04:12:00 | 19.921 | 0.016 | 2018 Jul 12 | 06:33:04 | 20.122 | 0.078 |
| 2018 Jul 11 | 03:24:08 | 19.863 | 0.016 | 2018 Jul 11 | 04:14:42 | 19.944 | 0.016 | 2018 Jul 12 | 06:34:14 | 20.066 | 0.076 |
| 2018 Jul 11 | 03:26:20 | 19.868 | 0.016 | 2018 Jul 11 | 04:15:53 | 19.952 | 0.017 | 2018 Jul 12 | 07:12:20 | 20.304 | 0.087 |
| 2018 Jul 11 | 03:27:26 | 19.892 | 0.017 | 2018 Jul 11 | 04:16:59 | 19.982 | 0.017 | 2018 Jul 12 | 07:13:34 | 20.258 | 0.089 |
| 2018 Jul 11 | 03:28:32 | 19.899 | 0.018 | 2018 Jul 11 | 04:18:05 | 19.957 | 0.016 | 2018 Jul 12 | 07:14:49 | 20.234 | 0.087 |
| 2018 Jul 11 | 03:29:38 | 19.892 | 0.019 | 2018 Jul 11 | 04:19:10 | 19.933 | 0.018 | 2018 Jul 12 | 07:16:03 | 20.186 | 0.085 |
| 2018 Jul 11 | 03:30:58 | 19.891 | 0.018 | 2018 Jul 11 | 04:20:16 | 19.934 | 0.018 | 2018 Jul 12 | 07:17:17 | 20.226 | 0.086 |
| 2018 Jul 11 | 03:32:04 | 19.881 | 0.017 | 2018 Jul 11 | 04:25:46 | 20.002 | 0.018 | 2018 Jul 31 | 22:04:16 | 20.501 | 0.033 |
| 2018 Jul 11 | 03:33:09 | 19.839 | 0.017 | 2018 Jul 11 | 04:26:59 | 19.991 | 0.016 | 2018 Jul 31 | 22:06:32 | 20.498 | 0.034 |
| 2018 Jul 11 | 03:34:15 | 19.916 | 0.018 | 2018 Jul 11 | 04:28:05 | 19.991 | 0.018 | 2018 Jul 31 | 22:08:48 | 20.432 | 0.032 |
| 2018 Jul 11 | 03:35:21 | 19.861 | 0.016 | 2018 Jul 11 | 04:29:10 | 19.990 | 0.018 | 2018 Jul 31 | 22:11:05 | 20.422 | 0.032 |
| 2018 Jul 11 | 03:36:27 | 19.882 | 0.016 | 2018 Jul 11 | 04:30:16 | 19.996 | 0.018 | 2018 Jul 31 | 22:13:21 | 20.504 | 0.041 |
| 2018 Jul 11 | 03:37:33 | 19.851 | 0.015 | 2018 Jul 11 | 04:31:22 | 20.004 | 0.017 | 2018 Jul 31 | 22:15:40 | 20.433 | 0.030 |
| 2018 Jul 11 | 03:38:38 | 19.863 | 0.015 | 2018 Jul 11 | 04:32:28 | 20.062 | 0.019 | 2018 Jul 31 | 22:17:56 | 20.472 | 0.038 |
| 2018 Jul 11 | 03:39:44 | 19.896 | 0.017 | 2018 Jul 12 | 02:40:42 | 20.198 | 0.083 | 2018 Jul 31 | 22:20:13 | 20.368 | 0.029 |
| 2018 Jul 11 | 03:40:51 | 19.852 | 0.016 | 2018 Jul 12 | 02:41:56 | 20.144 | 0.079 | 2018 Jul 31 | 22:22:29 | 20.332 | 0.026 |
| 2018 Jul 11 | 03:42:13 | 19.872 | 0.016 | 2018 Jul 12 | 02:44:24 | 20.207 | 0.083 | 2018 Jul 31 | 22:24:46 | 20.386 | 0.032 |
| 2018 Jul 11 | 03:43:19 | 19.843 | 0.015 | 2018 Jul 12 | 02:45:38 | 20.098 | 0.076 | 2018 Jul 31 | 22:27:06 | 20.433 | 0.027 |
| 2018 Jul 11 | 03:44:25 | 19.878 | 0.016 | 2018 Jul 12 | 03:52:30 | 20.093 | 0.077 | 2018 Jul 31 | 22:29:23 | 20.424 | 0.031 |
| 2018 Jul 11 | 03:45:31 | 19.844 | 0.016 | 2018 Jul 12 | 03:53:44 | 20.137 | 0.081 | 2018 Jul 31 | 22:33:57 | 20.444 | 0.036 |
| 2018 Jul 11 | 03:46:37 | 19.889 | 0.016 | 2018 Jul 12 | 03:54:59 | 20.098 | 0.078 | 2018 Jul 31 | 22:36:14 | 20.446 | 0.035 |
| 2018 Jul 11 | 03:47:43 | 19.928 | 0.015 | 2018 Jul 12 | 03:56:12 | 20.124 | 0.081 | 2018 Jul 31 | 22:38:34 | 20.410 | 0.033 |
| 2018 Jul 11 | 03:48:48 | 19.903 | 0.016 | 2018 Jul 12 | 03:57:27 | 20.243 | 0.085 | 2018 Jul 31 | 22:40:52 | 20.376 | 0.032 |
| 2018 Jul 11 | 03:49:54 | 19.893 | 0.015 | 2018 Jul 12 | 04:34:23 | 19.969 | 0.075 | 2018 Jul 31 | 22:43:07 | 20.415 | 0.037 |
| 2018 Jul 11 | 03:51:00 | 19.898 | 0.015 | 2018 Jul 12 | 04:35:37 | 20.034 | 0.076 | 2018 Jul 31 | 22:45:25 | 20.408 | 0.036 |
| 2018 Jul 11 | 03:52:06 | 19.917 | 0.015 | 2018 Jul 12 | 04:36:51 | 20.098 | 0.078 | 2018 Jul 31 | 22:47:40 | 20.297 | 0.030 |
| 2018 Jul 11 | 03:54:42 | 19.937 | 0.016 | 2018 Jul 12 | 05:16:03 | 20.229 | 0.085 | 2018 Jul 31 | 22:50:01 | 20.409 | 0.035 |
| 2018 Jul 11 | 03:55:47 | 19.915 | 0.016 | 2018 Jul 12 | 05:17:17 | 20.244 | 0.085 | 2018 Jul 31 | 22:54:34 | 20.313 | 0.038 |
| 2018 Jul 11 | 03:56:53 | 19.916 | 0.017 | 2018 Jul 12 | 05:18:31 | 20.262 | 0.085 | 2018 Jul 31 | 22:56:51 | 20.407 | 0.031 |
| 2018 Jul 11 | 03:57:59 | 19.929 | 0.015 | 2018 Jul 12 | 05:19:45 | 20.230 | 0.083 | 2018 Jul 31 | 22:59:07 | 20.450 | 0.036 |
| 2018 Jul 11 | 03:59:05 | 19.934 | 0.016 | 2018 Jul 12 | 05:20:59 | 20.209 | 0.081 | 2018 Jul 31 | 23:01:23 | 20.344 | 0.038 |
| 2018 Jul 11 | 04:01:52 | 19.880 | 0.016 | 2018 Jul 12 | 05:48:13 | 20.124 | 0.077 | 2018 Jul 31 | 23:05:56 | 20.328 | 0.037 |
| 2018 Jul 11 | 04:02:58 | 19.926 | 0.015 | 2018 Jul 12 | 05:49:27 | 20.144 | 0.078 | 2018 Jul 31 | 23:08:13 | 20.372 | 0.046 |
| 2018 Jul 11 | 04:04:04 | 19.898 | 0.015 | 2018 Jul 12 | 05:50:41 | 20.072 | 0.075 | 2018 Jul 31 | 23:10:29 | 20.432 | 0.051 |
| 2018 Jul 11 | 04:05:10 | 19.944 | 0.017 | 2018 Jul 12 | 06:29:36 | 20.049 | 0.076 | 2018 Jul 31 | 23:12:46 | 20.400 | 0.037 |
| 2018 Jul 11 | 04:06:16 | 19.940 | 0.017 | 2018 Jul 12 | 06:30:46 | 20.087 | 0.077 | 2018 Jul 31 | 23:15:02 | 20.462 | 0.044 |
| 2018 Jul 11 | 04:09:18 | 19.889 | 0.017 | 2018 Jul 12 | 06:31:55 | 20.082 | 0.076 | 2018 Jul 31 | 23:17:19 | 20.382 | 0.043 |

Supplementary Table - Continued

| Date[a] | UT[b] | $m_r$[c] | $\sigma_r$[d] | Date[a] | UT[b] | $m_r$[c] | $\sigma_r$[d] | Date[a] | UT[b] | $m_r$[c] | $\sigma_r$[d] |
|---|---|---|---|---|---|---|---|---|---|---|---|
| 2018 Jul 31 | 23:21:52 | 20.237 | 0.042 | 2018 Aug 1 | 01:09:04 | 20.340 | 0.053 | 2018 Aug 1 | 23:24:12 | 20.368 | 0.027 |
| 2018 Jul 31 | 23:24:08 | 20.286 | 0.062 | 2018 Aug 1 | 01:15:54 | 20.344 | 0.047 | 2018 Aug 1 | 23:26:28 | 20.424 | 0.028 |
| 2018 Jul 31 | 23:26:25 | 20.339 | 0.057 | 2018 Aug 1 | 01:20:26 | 20.270 | 0.074 | 2018 Aug 1 | 23:28:45 | 20.356 | 0.029 |
| 2018 Jul 31 | 23:28:41 | 20.270 | 0.037 | 2018 Aug 1 | 01:29:32 | 20.210 | 0.060 | 2018 Aug 1 | 23:33:18 | 20.398 | 0.030 |
| 2018 Jul 31 | 23:33:14 | 20.356 | 0.030 | 2018 Aug 1 | 21:50:41 | 20.274 | 0.020 | 2018 Aug 1 | 23:35:34 | 20.338 | 0.031 |
| 2018 Jul 31 | 23:35:37 | 20.313 | 0.038 | 2018 Aug 1 | 21:52:58 | 20.237 | 0.020 | 2018 Aug 1 | 23:37:51 | 20.315 | 0.026 |
| 2018 Jul 31 | 23:37:53 | 20.419 | 0.033 | 2018 Aug 1 | 21:55:17 | 20.290 | 0.019 | 2018 Aug 1 | 23:40:07 | 20.376 | 0.033 |
| 2018 Jul 31 | 23:40:09 | 20.365 | 0.038 | 2018 Aug 1 | 21:57:33 | 20.327 | 0.021 | 2018 Aug 1 | 23:46:57 | 20.409 | 0.038 |
| 2018 Jul 31 | 23:42:26 | 20.345 | 0.033 | 2018 Aug 1 | 21:59:52 | 20.353 | 0.022 | 2018 Aug 1 | 23:51:30 | 20.314 | 0.032 |
| 2018 Jul 31 | 23:44:43 | 20.266 | 0.026 | 2018 Aug 1 | 22:02:09 | 20.341 | 0.022 | 2018 Aug 1 | 23:53:46 | 20.324 | 0.034 |
| 2018 Jul 31 | 23:46:59 | 20.395 | 0.036 | 2018 Aug 1 | 22:04:31 | 20.366 | 0.021 | 2018 Aug 1 | 23:56:03 | 20.419 | 0.036 |
| 2018 Jul 31 | 23:49:16 | 20.354 | 0.041 | 2018 Aug 1 | 22:06:47 | 20.426 | 0.022 | 2018 Aug 1 | 23:58:23 | 20.367 | 0.035 |
| 2018 Jul 31 | 23:51:32 | 20.381 | 0.039 | 2018 Aug 1 | 22:09:03 | 20.395 | 0.022 | 2018 Aug 2 | 00:00:40 | 20.345 | 0.028 |
| 2018 Jul 31 | 23:53:49 | 20.410 | 0.039 | 2018 Aug 1 | 22:11:20 | 20.383 | 0.023 | 2018 Aug 2 | 00:02:56 | 20.380 | 0.032 |
| 2018 Jul 31 | 23:58:22 | 20.481 | 0.042 | 2018 Aug 1 | 22:13:36 | 20.442 | 0.025 | 2018 Aug 2 | 00:05:13 | 20.405 | 0.030 |
| 2018 Aug 1 | 00:00:38 | 20.432 | 0.055 | 2018 Aug 1 | 22:18:09 | 20.479 | 0.026 | 2018 Aug 2 | 00:12:02 | 20.410 | 0.029 |
| 2018 Aug 1 | 00:02:54 | 20.488 | 0.060 | 2018 Aug 1 | 22:24:59 | 20.493 | 0.025 | 2018 Aug 2 | 00:14:19 | 20.449 | 0.034 |
| 2018 Aug 1 | 00:05:11 | 20.511 | 0.044 | 2018 Aug 1 | 22:27:15 | 20.524 | 0.025 | 2018 Aug 2 | 00:16:35 | 20.415 | 0.032 |
| 2018 Aug 1 | 00:07:27 | 20.467 | 0.038 | 2018 Aug 1 | 22:29:32 | 20.510 | 0.025 | 2018 Aug 2 | 00:18:52 | 20.361 | 0.031 |
| 2018 Aug 1 | 00:09:44 | 20.565 | 0.050 | 2018 Aug 1 | 22:31:48 | 20.524 | 0.027 | 2018 Aug 2 | 00:21:08 | 20.430 | 0.031 |
| 2018 Aug 1 | 00:12:00 | 20.549 | 0.043 | 2018 Aug 1 | 22:36:21 | 20.508 | 0.026 | 2018 Aug 2 | 00:23:25 | 20.344 | 0.032 |
| 2018 Aug 1 | 00:21:11 | 20.510 | 0.036 | 2018 Aug 1 | 22:38:38 | 20.526 | 0.024 | 2018 Aug 2 | 00:25:41 | 20.329 | 0.032 |
| 2018 Aug 1 | 00:23:28 | 20.618 | 0.053 | 2018 Aug 1 | 22:40:54 | 20.520 | 0.025 | 2018 Aug 2 | 00:27:58 | 20.372 | 0.031 |
| 2018 Aug 1 | 00:25:44 | 20.585 | 0.050 | 2018 Aug 1 | 22:45:27 | 20.520 | 0.026 | 2018 Aug 2 | 00:30:14 | 20.432 | 0.035 |
| 2018 Aug 1 | 00:28:01 | 20.491 | 0.051 | 2018 Aug 1 | 22:52:17 | 20.507 | 0.024 | 2018 Aug 2 | 00:34:47 | 20.479 | 0.033 |
| 2018 Aug 1 | 00:30:17 | 20.549 | 0.045 | 2018 Aug 1 | 22:54:33 | 20.374 | 0.023 | 2018 Aug 2 | 00:37:03 | 20.533 | 0.031 |
| 2018 Aug 1 | 00:32:33 | 20.562 | 0.053 | 2018 Aug 1 | 22:56:49 | 20.384 | 0.022 | 2018 Aug 2 | 00:39:20 | 20.489 | 0.032 |
| 2018 Aug 1 | 00:34:50 | 20.620 | 0.052 | 2018 Aug 1 | 22:59:06 | 20.394 | 0.023 | 2018 Aug 2 | 00:41:36 | 20.532 | 0.036 |
| 2018 Aug 1 | 00:43:56 | 20.545 | 0.050 | 2018 Aug 1 | 23:01:27 | 20.477 | 0.025 | 2018 Aug 2 | 00:57:37 | 20.643 | 0.035 |
| 2018 Aug 1 | 00:46:12 | 20.432 | 0.058 | 2018 Aug 1 | 23:03:44 | 20.485 | 0.027 | 2018 Aug 2 | 00:59:54 | 20.683 | 0.042 |
| 2018 Aug 1 | 00:48:29 | 20.361 | 0.046 | 2018 Aug 1 | 23:06:00 | 20.489 | 0.026 | 2018 Aug 2 | 01:02:10 | 20.635 | 0.035 |
| 2018 Aug 1 | 00:50:45 | 20.325 | 0.051 | 2018 Aug 1 | 23:08:17 | 20.386 | 0.026 | 2018 Aug 2 | 01:04:27 | 20.562 | 0.034 |
| 2018 Aug 1 | 00:53:02 | 20.330 | 0.043 | 2018 Aug 1 | 23:10:33 | 20.379 | 0.027 | 2018 Aug 2 | 01:06:43 | 20.613 | 0.040 |
| 2018 Aug 1 | 00:57:35 | 20.285 | 0.041 | 2018 Aug 1 | 23:12:50 | 20.384 | 0.026 | 2018 Aug 2 | 01:09:01 | 20.538 | 0.041 |
| 2018 Aug 1 | 00:59:51 | 20.252 | 0.046 | 2018 Aug 1 | 23:15:06 | 20.418 | 0.028 | 2018 Aug 2 | 01:11:21 | 20.521 | 0.037 |
| 2018 Aug 1 | 01:04:24 | 20.292 | 0.065 | 2018 Aug 1 | 23:17:23 | 20.421 | 0.028 | 2018 Aug 2 | 01:13:42 | 20.538 | 0.038 |
| 2018 Aug 1 | 01:06:48 | 20.314 | 0.051 | 2018 Aug 1 | 23:21:55 | 20.316 | 0.024 | 2018 Aug 2 | 01:16:02 | 20.529 | 0.035 |

Supplementary Table - Continued

| Date[a] | UT[b] | $m_{r'}$[c] | $\sigma_{r'}$[d] |
|---|---|---|---|
| 2018 Aug 2 | 01:18:24 | 20.572 | 0.037 |
| 2018 Aug 2 | 01:20:46 | 20.546 | 0.035 |
| 2018 Aug 2 | 01:25:23 | 20.486 | 0.037 |
| 2018 Aug 2 | 01:27:42 | 20.447 | 0.039 |
| 2018 Aug 2 | 01:30:03 | 20.484 | 0.037 |
| 2018 Aug 2 | 21:47:06 | 20.558 | 0.077 |
| 2018 Aug 2 | 21:49:23 | 20.509 | 0.075 |
| 2018 Aug 2 | 21:51:39 | 20.439 | 0.073 |
| 2018 Aug 2 | 21:53:56 | 20.479 | 0.074 |
| 2018 Aug 2 | 21:56:12 | 20.448 | 0.073 |
| 2018 Aug 2 | 21:58:29 | 20.470 | 0.074 |
| 2018 Aug 2 | 22:00:45 | 20.417 | 0.072 |
| 2018 Aug 2 | 22:03:02 | 20.402 | 0.071 |
| 2018 Aug 2 | 22:05:18 | 20.435 | 0.073 |
| 2018 Aug 2 | 22:07:35 | 20.430 | 0.072 |
| 2018 Aug 2 | 22:09:51 | 20.405 | 0.072 |
| 2018 Aug 2 | 22:12:08 | 20.398 | 0.071 |
| 2018 Aug 2 | 22:14:24 | 20.359 | 0.070 |
| 2018 Aug 2 | 22:16:41 | 20.365 | 0.070 |
| 2018 Aug 2 | 22:18:57 | 20.359 | 0.070 |
| 2018 Aug 2 | 22:21:14 | 20.363 | 0.070 |
| 2018 Aug 2 | 22:23:30 | 20.361 | 0.070 |
| 2018 Aug 2 | 22:25:47 | 20.391 | 0.071 |
| 2018 Aug 2 | 22:28:03 | 20.439 | 0.073 |
| 2018 Aug 2 | 22:30:29 | 20.410 | 0.072 |
| 2018 Aug 2 | 22:32:45 | 20.433 | 0.072 |
| 2018 Aug 2 | 22:35:02 | 20.468 | 0.073 |
| 2018 Aug 2 | 22:37:18 | 20.459 | 0.073 |
| 2018 Aug 2 | 22:39:35 | 20.496 | 0.074 |
| 2018 Aug 2 | 22:41:51 | 20.511 | 0.075 |
| 2018 Aug 2 | 22:44:14 | 20.522 | 0.075 |
| 2018 Aug 2 | 22:46:30 | 20.564 | 0.077 |
| 2018 Aug 2 | 22:48:47 | 20.607 | 0.079 |
| 2018 Aug 2 | 22:51:03 | 20.560 | 0.077 |

Notes.
[a] UT date of observation.
[b] UT at midpoint of the exposure.
[c] Observed $r'$-band magnitude.
[d] Statistical uncertainty in $r'$-band magnitude.